\documentclass[12pt]{article}
\usepackage{color}
\usepackage{epsfig}
\usepackage{amssymb,latexsym,amsmath}

    \def\ee{\end{equation}}
    \def\ea{\end{eqnarray}}

\begin{document}
\title{\textbf{Global stability analysis for cosmological models with nonminimally coupled scalar fields}}
\author{Maria A. Skugoreva$^{1}\footnote{masha-sk@mail.ru}$, Alexey V. Toporensky$^{2,3}$\footnote{atopor@rambler.ru}, and Sergey Yu. Vernov$^{4}$\footnote{svernov@theory.sinp.msu.ru}\vspace*{3mm} \\
\small $^{1}$Peoples Friendship University of Russia, Moscow, 117198, Russia\\
\small $^{2}$Sternberg Astronomical Institute, Lomonosov Moscow State University,\\
\small Moscow, 119991, Russia\\
\small $^{3}$Kazan Federal University, Kremlevskaya 18, Kazan, 420008, Russia,\\
\small $^{4}$Skobeltsyn Institute of Nuclear Physics, Lomonosov Moscow State University, \\
\small  Moscow, 119991,  Russia}

\date{ \ }
\maketitle
\begin{abstract}
  We explore dynamics of cosmological models with a nonminimally coupled scalar field evolving on a spatially flat Friedmann--Lema\^{i}tre--Robertson--Walker background. We consider cosmological models including the Hilbert--Einstein curvature term and the $N$ degree monomial of the scalar field nonminimally coupled to gravity. The potential of the scalar field is the $n$ degree monomial or polynomial.
We describe several qualitatively different types of dynamics depending on values of power indices $N$ and $n$.
 We identify that three main possible pictures correspond to $n<N$, $N<n<2N$ and $n>2N$ cases. Some special features connected with the important cases of $N=n$ (including the quadratic potential with quadratic coupling) and $n=2N$ (which shares its asymptotics with the potential of the Higgs-driven inflation) are described separately.
A global qualitative analysis allows us to cover the most interesting cases of small $N$ and $n$ by a limiting number of phase-space diagrams. The influence of the cosmological constant to the global features of dynamics is also studied.
\end{abstract}
\maketitle

\section{Introduction}

The assumption that General Relativity is the
correct theory of gravity leads to the conclusion that the observable data~\cite{cosmo-obser,PlanckInflation,Planck2013} give the strong support that there exists and currently dominates a smoothly distributed, slowly varying cosmic fluid with negative pressure,  so-called dark energy~\cite{DE_rev,DINDE}.
The simplest way to describe the dark energy is to add the cosmological constant to the  Einstein--Hilbert action.
Another popular variant is to consider models with scalar fields~\cite{DINDE,Tsujikawa:2013fta}.
Models with scalar fields are very useful to describe the observable evolution of the Universe as the dynamics of the spatially flat Friedmann--Lema\^{i}tre--Robertson--Walker (FLRW) background and cosmological perturbations.
That is why scalar fields play an essential role in modern cosmology.

The models with the Ricci scalar multiplied by a function of the scalar field are being intensively studied in cosmology~\cite{induced,nonmin-inf,Kaiser1994,HiggsInflation,Cooper:1982du,KKhT,
Elizalde,Cerioni,Kamenshchik:2012rs,Gannouji:2006jm,Szydlo,
KTV2011,Sami:2012uh,ABGV,KTVV2013,KPTVV2013} (see also~\cite{Book-Capozziello-Faraoni,Fujii_Maeda,NO-rev} and references therein).
Induced gravity models, wherein the curvature arises as a quantum effect~\cite{Sakharov}, and models, where both the Hilbert--Einstein term and the  scalar field squared multiplied by the scalar curvature are present, have been applied to quantum cosmology~\cite{nonmin-quant} and are being intensively studied in the inflationary cosmology~\cite{induced,nonmin-inf,Kaiser1994,HiggsInflation}.
Note that predictions of the simplest inflationary models with minimally coupling scalar fields are in disagreement with the Planck2013
results~\cite{Planck2013} and some of these inflationary scenarios have been improved by adding a tiny nonminimal coupling of the inflaton field to gravity~\cite{GB2013,KL2013}. The Higgs-driven inflation has attracted a lot of attention~\cite{HiggsInflation}. The recent discovery of the Higgs boson~\cite{discovery} makes this model especially attractive. The predictions of this inflation model is very
close to predictions of the Starobinsky inflation~\cite{Starobinsky} (see~\cite{BezrukovR2Higgs} for details).

The number of integrable cosmological models based on scalar fields is rather limited. The most popular integrable cosmological model is the model with a minimally coupled scalar field and a self-interaction exponential potential~\cite{Lucchin,gen-exp}. In~\cite{Fre}, the general classification of integrable cosmological models based on scalar fields was suggested and studied in great detail. Integrable models with nonminimally coupled scalar fields have been found in~\cite{KPTVV2013}.

In this situation it is reasonable to search for asymptotic regimes in the theory under consideration.
Note that dynamical system  methods are extensively used for analysis of cosmological models with scalar fields~\cite{Szydlo,Sami:2012uh,CapozzelloPhasespaceview,FaraoniDS,GalileonDS,Leon}.
Using these methods several different asymptotic regimes have been found, and their stability has been investigated for the simplest case of quadratic coupling in \cite{CapozzelloPhasespaceview} and for other power-law coupling in \cite{Sami:2012uh}. However, as such analysis usually requires transition to expansion normalized variables which can be not smooth in some points, this results in the fact that some solutions can be lost in this procedure. That is why a global analysis of the dynamics in question is an important counterpart to the description of locally stable regimes -- apart from the fact that phase-space diagrams are very useful in visualizing the dynamical picture (especially for the considered two-dimensional problem), we also can be sure that we did not miss some important regimes  during local analysis.  Note that the phase-space diagrams are actively used to analyze dynamics of a cosmological model with scalar fields (maybe with nonstandard kinetic term) minimally coupled to gravity~\cite{PhantomPhPort,FaraoniMC}. A combination of the phase-space and stability analysis is a systematic way to explore the possible
cosmological behaviors.

Recently a set of phase diagrams for the theory with a quadratic coupling have been constructed
in \cite{CapozzelloPhasespaceview,Szydlo}. In our paper, we consider a
more general case. We investigate the dynamics of cosmological models including the Hilbert--Einstein curvature term, a monomial function of the scalar field coupled to gravity and polynomial potentials.

Using the analysis of \cite{Sami:2012uh}, it is possible to show that for the case of $\xi<0$
(the only case we consider in the present paper) the form of phase diagrams does not
 depend on $\xi$ and depends only upon relations between power indexes of the coupling function and the scalar field monomial potential. This allows us to characterize completely the global feature of
 cosmological dynamics in the model in question
using a limiting number of phase portraits. We start with listing known asymptotic
 regimes from \cite{CapozzelloPhasespaceview, Sami:2012uh} and
show how they are incorporated into a global picture by constructing phase-space diagrams.

The paper is organized as follows.
In Section~2, we examine the Friedmann equations for models with a nonminimally coupled scalar field. In Section~3 we consider de Sitter solutions and analyze their stability with the help of the effective potential.
In Section~4 we consider asymptotic solutions for the considering model. We also present Rizmaikin-type~\cite{Ruzmaikin} solutions and the corresponding potentials. In Sections~5 and 6 we use the phase-space diagrams for the global qualitative analysis of the cosmological dynamics.
In Section~5 we consider monomial potentials, and more complicated potentials are considered in Section~6. Section~7 is devoted to the conclusions.

\section{Basic equations}
In this paper, we consider models described by the following action:
\begin{equation}
\label{action}
S=\int d^4 x \sqrt{-g}\left[ \frac12\left(m_{p}^2-\xi B(\varphi)\right)R-\frac12g^{\mu\nu}\varphi_{,\mu}\varphi_{,\nu}-V(\varphi)\right],
\end{equation}
where $m_{p}^2$ is a constant that corresponds to the Planck mass in the Einstein gravity, $\xi$ is the dimensionless coupling constant, $B(\varphi)$ and $V(\varphi)$ are differentiable functions of the scalar field~$\varphi$. We use the
signature $(-,+,+,+)$, and~$g$ is the determinant of the metric tensor
$g_{\mu\nu}$.

Consider the evolution of a homogeneous scalar field on a spatially
flat FLRW universe with the interval
\begin{equation*}
ds^2={}-dt^2+a^2(t)\left(dx_1^2+dx_2^2+dx_3^2\right).
\end{equation*}
In this metric the Einstein equations derived from the variation of action (\ref{action}) are as follows:
\begin{equation}
\label{equ00}
H^2=\frac{1}{3m_{p}^2}\left(\frac{1}{2}{\dot
{\varphi}}^{2}+V+3\xi\left(H  B'\dot {\varphi}+H^{2}B\right)
\right),
\end{equation}
\begin{equation}
\label{equ11}
\left(m_{p}^2-\xi B\right)\left[2\dot H+3H^2\right]={}-\frac{1}{2}{\dot
{\varphi}}^{2}+V+2\xi H  B'\dot {\varphi}+\xi\left(B''\dot{\varphi}^2+B'\ddot{\varphi}\right),
\end{equation}
where differentiation with respect to time $t$ is denoted by a dot, the prime indicates the derivative with respect to the scalar field
$\varphi$, and the Hubble parameter is  the logarithmic derivative of the scale factor:
$H=\dot a/a$.

 The variation of action (\ref{action}) with respect to $\varphi$ gives
the Klein--Gordon equation
\begin{equation}
\label{equ_sigma}
\ddot {\varphi}+3 H \dot {\varphi}+\frac{1}{2}\xi B'R+V'=0,
\end{equation}
and the Ricci scalar in
the FLRW metric is given by $R=6\left(2H^2+ \dot {H}\right)$. It is easy to show that
\begin{equation}
\label{equR}
\left(m_p^2-\xi B\right)R= {}-{\dot {\varphi}}^{2} +4
V(\varphi)+3\xi\left(3 H  B'\dot{\varphi}  + B''{{\dot
\varphi}}^{2}+B'\ddot {\varphi} \right).
\end{equation}

Combining equations (\ref{equ00}) and (\ref{equ11}) we get
\begin{equation}
\label{Fr21}
2\left(m_p^2-\xi B\right)\dot H+\xi \dot B H-\xi\ddot B +\dot\varphi^2=0.
\end{equation}

 For convenience,
we shall use the system of units with $m_p^2=1/6$ (the same units have been used in~\cite{Sami:2012uh}).
From the system of equations (\ref{equ00}), (\ref{equ11}), and (\ref{equ_sigma}) we get the following system of the first order equations:
\begin{equation}
\label{eq_Ht}
 \dot H =\frac{3{\psi}^2(\xi B''-1)-3\xi B'\left(4H \psi+V'+6\xi B' H^2\right)}{1-6\xi B+9\xi^2{B'}^2},
\end{equation}
\begin{equation}
\label{eq_phitt}
\begin{split}
\dot\varphi&=\psi,\\
\dot{\psi}&={}-3H \psi+\frac{3\xi B'(1-3\xi B'')}{1-6\xi B+9{\xi}^2{B'}^2}\psi^2-\frac{V'+6\xi(2V B' -V' B)}{1-6\xi B+9{\xi}^2{B'}^2}.
\end{split}
\end{equation}

The Hubble parameter $H$ can be found from the quadratic equation~(\ref{equ00}),
therefore,
\begin{equation}
\label{H}
H=\frac{3\xi\dot {\varphi} B'\pm\sqrt{9\xi^2{\dot {\varphi}}^2 B^{'2}+(1-6\xi B)(\dot{\varphi}^{2}+2 V)}}{1-6\xi B}.
\end{equation}

Note that $H$ is a continuous function, so, if the radicand is not equal to zero at any moment of time, then the initial conditions define the sign "$+$" or "$-$" in (\ref{H}). For example, if $\xi<0$ and $B\geqslant 0$, then for a positive-definite potential $V(\varphi)$ this sign is defined uniquely.

In this paper, we choose the sing "$+$", substitute the corresponding Hubble parameter into~(\ref{eq_phitt}), and get that the second order system~(\ref{eq_phitt}) determines the dynamics of the considering cosmological model.
Our discussion will be restricted to power-law functions $B(\varphi)=\varphi^N$. Most of our discussions deal with power-law potentials $V(\varphi)=\varphi^n$, certain particular cases of more general potentials $V=V_0\varphi^n+\lambda\varphi^{n_1}$ will be considered as well. In this paper, we study only positively defined potentials, $\xi<0$ and coupling functions with $N\geqslant 2$ and $n\geqslant 2$, $n_1\geqslant 0$ that are chosen to be even numbers.

\section{Stability analysis with the effective potential}

For qualitative description of the scalar field evolution it is useful to find points where the driving force vanishes, so $\varphi=\mathrm{const}$ is a solution.
One can see from Eq.~(\ref{eq_phitt}) that such points correspond to critical points of the effective potential
\begin{equation*}
V_{eff}=\frac{V(\varphi)}{(1-6\xi B(\varphi))^2},
\end{equation*}
because
\begin{equation*}
V'_{eff}=\frac{V'+6\xi(2V B' -V' B)}{(1-6\xi B(\varphi))^3}.
\end{equation*}

 After appropriate transformation of the scalar field this potential becomes proportional to the potential in the Einstein frame ---  the theory conformally invariant to the initial one, where a scalar field
is minimally coupled to gravity. We do not provide full transition to the Einstein frame here because  the form of the effective potential given above without any additional transformation of the scalar field gives us  enough information about the
stability of de Sitter solutions in the theory under consideration.  It is evident that the form of the effective potential having its extreme
at $\varphi=\mathrm{const}$ as a solution is not unique, so  we use the form chosen for its simplicity (it was used already in the similar
context in Ref.\cite{KKhT}).  The important thing is that the third term in the right-hand side of Eq.~(\ref{eq_phitt}) differs from $V_{eff}'$ by
function $(1-6\xi B(\varphi))$ which is always positive (for $\xi<0$) and thus does not affect global properties of the dynamics.

Let us rewrite equations (\ref{equ00}) and (\ref{Fr21})
in the form similar to the Friedmann equations in the Einstein frame. We introduce a new variable
\begin{equation*}
P\equiv \frac{H}{\sqrt{U}}+\frac{U'\dot\varphi}{2U\sqrt{U}},
\end{equation*}
where $U(\varphi)=\frac12\left(\frac{1}{6}-\xi B(\varphi)\right)$.
In terms of $P$ Eq.~(\ref{equ00}) has the following form
\begin{equation}
\label{equP}
3P^2=\frac{U+3{U'}^2}{4U^3}{\dot\varphi}^2+\frac{V}{2U^2}=A{\dot\varphi}^2+72V_{eff},
\end{equation}
where $A\equiv (U+3{U'}^2)/(4U^3)$. We consider $U(\varphi)>0$ only, so $A(\varphi)>0$ at any $\varphi$.
Using (\ref{equ00}), we get from Eq.~(\ref{Fr21}) the following equation:
\begin{equation}
\label{Fr21Qm}
\dot P={}-A\sqrt{U}\,{\dot\varphi}^2.
\end{equation}
Now we differentiate (\ref{equP}) over time, substitute (\ref{Fr21Qm}) and get
\begin{equation}
\begin{split}
\label{system2}
\dot\varphi&=\psi\,,\\
\dot\psi&={}-3P\sqrt{U}\psi-\frac{A'}{2A}\psi^2-36\frac{V'_{eff}}{A}.
\end{split}
\end{equation}

The de Sitter solutions corresponds to $\psi=0$, and hence,
$V'_{eff}(\varphi_{dS})=0$, in other words
\begin{equation*}
V'(\varphi_{dS})U(\varphi_{dS})=2V(\varphi_{dS})U'(\varphi_{dS}).
\end{equation*}
The corresponding Hubble parameter is
\begin{equation}
\label{Hf}
H_{dS}=P_{dS}\sqrt{U(\varphi_{dS})}
=\pm\sqrt{\frac{V(\varphi_{dS})}{6U(\varphi_{dS})}}
=\pm\sqrt{\frac{V'(\varphi_{dS})}{12U'(\varphi_{dS})}}\,.
\end{equation}

Let us consider the Lyapunov stability of a de Sitter solution. Substituting
\begin{equation}
\varphi(t)=\varphi_{dS}+\varphi_1(t),\qquad \psi(t)=\psi_1(t),
\end{equation}
into (\ref{system2}), we get the following linear system on $\varphi_1(t)$ and $\psi_1(t)$:
\begin{equation}
\label{linsystem}
\begin{split}
\dot\varphi_1&=\psi_1,\\
\dot\psi_1&={}-36\frac{V''_{eff}(\varphi_{dS})}{A}\varphi_1-3H_{dS}\psi_1.
\end{split}
\end{equation}

So, the considering de Sitter solution is stable under conditions $H_{dS}>0$ and $V''_{eff}(\varphi_{dS})>0$. In other words, the model has a stable de Sitter solution only if the potential $V_{eff}$ have a minimum. Note that this conclusion is valid for  arbitrary differentiable functions $U$ and $V$, under the condition $U(\varphi_{dS})>0$.

In the case of power-law potential $V(\varphi)=\varphi^n$ there exists the following de Sitter solution:
\begin{equation}
\label{dSSol}
\begin{array}{l}
a(t)=a_0 e^{H_{dS}(t-t_0)},\\
H_{dS}=\displaystyle \pm\sqrt{-\frac{V_0 n \varphi_{dS}^{n-N}}{6\xi N}},\\
\varphi_{dS}=\displaystyle \pm\left[\frac{n}{6\xi (n-2 N)}\right]^{1/N}.
\end{array}
\end{equation}
Note that this solution does not exist for $n\geqslant2N$. So, we consider the case $n<2N$. Also, we assume that the potential is positive: $V_0>0$.
The stability of de Sitter solutions in the case of power-law potentials has been analyzed in~\cite{Sami:2012uh}. The use of the effective potential makes this analysis more trivial.
Indeed, in the case of a power-law potential $V$ the de Sitter solution is located at a point of a local maximum of the effective potential $V_{eff}$ and is unstable at  $\xi<0$ and any values of $N$ and $n$. For example, at $n=2$ we get
\begin{equation*}
V''_{eff}(\varphi_{dS})={}-\frac{2V_0(N-1)^3}{N^2}<0.
\end{equation*}

Also, there are stable Minkowski solutions at the point $\varphi=0$.
Note that potentials $V$ having more complicated form than a simple monomial can generate a stable de Sitter solution even without an explicit $\Lambda$ term in the action (see the end of the Section~\ref{SectionPolyPot}, where the corresponding effective potential is presented in Fig.~\ref{V6_2}).

\section{Asymptotic solutions}
In this section we consider regimes for which $\varphi \to \pm \infty$.  Though the effective potential can not help in identifying all of them,
it can be useful in order to find some sort of stable solutions. The reason is that the first and second terms in the right-hand side of Eq.~(\ref{eq_phitt})
represent in the case of
negative $\xi$
the positive friction terms if $\varphi \to \infty$ -- though the second term can not now be  considered as small, it is  negative in this limit
as well as the first term.
 As a result,
   system~(\ref{eq_phitt}) describes behavior of an oscillator with a driving force (the third term in the right-hand side of (\ref{eq_phitt})) and a positive friction (the first and second
terms). This means that influence of initial conditions should be washed out after some time, and the dynamics appears to be determined
by the driving force solely. This driving force pushes the field $\varphi$ to infinity if the effective potential is decreasing as a function
of $\varphi$ for large $\varphi$. Such behavior of the effective potential occurs for certain important cases (see below).
In the next two subsections we present these regimes explicitly, as well as other regimes which can not
be guessed from the form of the effective potential (all of them appear to be unstable).

\subsection{The case of $N>2$.}

Now we list asymptotic solutions for system (\ref{eq_Ht})--(\ref{eq_phitt}) in the limit $\varphi \to \infty$ that has been obtained in~\cite{Sami:2012uh}. We remind these solutions  and show the correspondence between behaviors of these solutions and properties of the  effective potentials. These solutions do not require the power index of the potential to be integer, though we
will not consider a noninteger (as well as odd integer) $n$ in the present paper.  As the case of $N=2$ is exceptional, so, we consider it separately in the next subsection.\\
\textbf{1.}  Consistency analysis shows that the solution
\begin{equation}
\begin{array}{l}
\label{9}
a(t)=a_0{|t-t_0|}^{1/2},\\
\varphi(t)={\varphi}_0{|t-t_0|}^{{}-1/(2N)}.
\end{array}
\end{equation}
exists for $N>2$, $n<5N$, $t\rightarrow t_0$. This solution  corresponds to the situation  in which the scalar field potential is negligible. As this regime is an unstable node  \cite{Sami:2012uh}, it represents a source point  on phase-space diagrams. It is interesting that the corresponding
regime for a massless minimally coupled scalar field  has a scale factor changing as time in the power $1/3$ instead of $1/2$ here, so the
$N>2$ case represents a clear discontinuity in the limit $\xi \to 0$.

\textbf{2.} The power-law solution\footnote{Note that  $\varphi\rightarrow\infty$ corresponds to $t\rightarrow\infty$ at $n<N$ and $t\rightarrow t_0$ at $n>N$.}
\begin{equation}
\begin{array}{l}
a(t)=a_0{|t-t_0|}^{\frac{(N+n)N}{(2 N-n)(N-n)}},\\
\varphi(t)={\varphi}_0{|t-t_0|}^{2/(N-n)},
\label{powerlawN}
\end{array}
\end{equation}
is stable for $n<2N$ and unstable for $n>2N$. This solution (in contrast to the previous one) depends on the scalar field  potential.
The effective potential is decreasing for large $\varphi$ if $n<2N$, so stability of this regime can be  qualitatively explained by arguments
presented in the beginning of this section.
  As for the particular properties of instability for $n>2N$, the full analysis gives us the following
result~\cite{Sami:2012uh}: for very steep potentials with $n>5N$ it is an unstable node (and plays the role of source point when the point
corresponding to solution (\ref{9}) does not exist), for $2N<n<5N$ it is a saddle.
It is clear that this solution is absent for $n=N$ or $n=2N$. For the former case a special stable regime exists described below.\\
\textbf{3}. In the case of $N=n$ there exists the following solution with a constant Hubble parameter $H=H_0$:
\begin{equation}
\begin{array}{l}
a(t)=a_0 e^{H_0(t-t_0)},\\
\varphi(t)=\varphi_0 e^{H_0(t-t_0)/N},
\end{array}
\end{equation}
where ${H_0}^2={}-V_0/(6\xi)$. Note that the scalar field $\varphi$
increases exponentially giving rise to exponentially decreasing
the effective Newtonian gravitational constant. For this reason such a solution does not qualify as a de Sitter one. We keep the notion
of a de Sitter solution for the solution with $H=\mathrm{const}$ and $\varphi=\mathrm{const}$.  The case of $n=2N$ will be discussed in Subsection 4.3.

\subsection{The case of $N=2$.}

\textbf{1.} There exists the following effective massless solution:
\begin{equation}
\begin{array}{l}
a(t)=a_0{|t-t_0|}^{\frac{1}{3-12 \xi-2 \sqrt{6 \xi(6 \xi-1)} }},\\
\varphi(t)={\varphi}_0{|t-t_0|}^{\frac{-6 \xi-\sqrt{6 \xi(6
\xi-1)}}{3-12 \xi-2 \sqrt{6 \xi(6  \xi-1)}}}.
\end{array}
\label{N2sol1}
\end{equation}
 In contrast to its $N \ne 2$ analog, given by (\ref{9}), the power indices of solution (\ref{N2sol1}) do depend on $\xi$. However, for
$\xi<0$ and $n \leqslant 10$ the stability analysis shows that
properties of this point do not "feel" $\xi$ -- it is always a source on phase space diagrams.
For steep potentials ($n>10$) this solution exists only in an interval $\xi_{cr}<\xi<0$ where $\xi_{cr}=-\frac{6}{(n+2)(n-10)}$ (in the $\xi>0$ case the situation
is more complicated, and we will not consider it in the present paper).

\textbf{2.} There is an analog of the power-law solution written down above for general $N$:
\begin{equation}
\begin{array}{l}
a(t)=a_0{|t-t_0|}^{\frac{2 (\xi(2+n)-1)}{\xi(n-2)(n-4)}},\\
\varphi(t)={\varphi}_0{|t-t_0|}^{2/(2-n)}
\end{array}
\end{equation}
From the form of the effective potential we can see
that it is an attractor for $n<4$. This solution has been found more than two decades ago in \cite{old}, and its
 Big Rip nature  for $2<n<4$ has been discussed in \cite{Gannouji:2006jm}.
Using methods of \cite{CapozzelloPhasespaceview, Sami:2012uh}, it is possible to show that this regime is  a saddle point on the phase-space diagram for $4<n \leqslant 10$. Moreover,
if $n>10$ it becomes an unstable node for $\xi< \xi_{cr}$ -- exactly in  the range where solution (12) does not exist.
As for the general case, we have a special form
for the case of equal power indices.

\textbf{3.} In the case of $n=2$ there exists the following solution with a constant Hubble parameter $H=H_0$:
\begin{equation}
\begin{array}{l}
a(t)=a_0 e^{H_0(t-t_0)},\\
\varphi(t)=\varphi_0 e^{\frac{2 H_0\xi(t-t_0)}{4\xi-1}},
\end{array}
\label{18}
\end{equation}
where ${H_0}^2={}-\frac{V_0{(4\xi-1)}^2}{\xi(96\xi^2-34\xi+3)}$. As its $N>2$ analog, this regime is an attractor.

The case of $n=4$ will be considered in the next subsection.

\subsection{Analog of Ruzmaikin solution}
In this subsection we consider a solution which shares the time behavior of the Hubble parameter with the known Ruzmaikin solution in $R+R^2$ gravity~\cite{Ruzmaikin}. Since this solution has not been found in \cite{Sami:2012uh}, we consider it here in more detail.

We assume that there is an asymptotic solution in the form $H=H_0 \tau$, $\varphi=\varphi_0 {\tau}^{\alpha}$, where $\tau=t-t_0$, $t_0$ is a constant. Substituting this solution into Eq.~(\ref{Fr21}), we
get
\begin{equation*}
\frac{H_0}{3}+\left(N\alpha-2\right)\xi \varphi_0^NH_0 {\tau}^{\alpha N}-
\xi N\alpha(\alpha-1)\varphi_0^N{\tau}^{\alpha N-2}-\xi N(N-1){\alpha}^2\varphi_0^N{\tau}^{\alpha N-2} +{\alpha}^2\varphi_0^2{\tau}^{2\alpha-2}=0.
\end{equation*}
In the limit $t\rightarrow\infty$ we neglect the smallest term ${\alpha}^2\varphi_0^2{\tau}^{2\alpha-2}$ and equate terms with identical power indexes. Analyzing the terms proportional to $\tau^{\alpha N}$ we get the power index for the scalar field
$\alpha=2/N$. Other terms give the equation connecting $H_0$ and $\varphi_0$:
\begin{equation}
H_0=6\varphi_0^N\xi.
\end{equation}
Now we substitute $H=H_0 \tau$, $\varphi=\varphi_0 \tau^{2/N}$ in (\ref{equ00}) and get
\begin{equation}
V(\varphi)=\frac{1}{2}\left[H^2(1-6\xi B)-{\dot{\varphi}}^{2}-6\xi H \dot B\right]=-108\xi^3\varphi_0^N\varphi^{2N}-18\xi^2\varphi_0^N\varphi^{N}-\frac{2}{N^2 \varphi_0^N}\varphi^{2-N}.
\end{equation}
Thereby, we get that the potential has the form
\begin{equation*}
V=V_0\varphi^{2N}+\frac{V_0}{6\xi}\varphi^{N}+\dots
\end{equation*}
and the Ruzmaikin-type solution has the following numeric parameters:
\begin{equation}
\varphi_0^N={}-\frac{V_0}{108\xi},\qquad H_0^2=\frac{V_0^2}{324\xi^2}.
\end{equation}

The first term in the potential gives the desired asymptotic for $\varphi \to \infty$ if $n=2N$.

The particular case of $N=2$ and $n=4$ is well known because it is used to describe the Higgs-driven inflation. The corresponding
effective potential is asymptotically flat (this is valid also for a general $n=2N$ case), as well as the potential in the Einstein frame.
The latter is the reason why inflation does not need extremely small self-coupling coefficient as in the minimally coupled field case.
A connection between this model and modified gravity has already been  remarked. In particular, the theory in question shares the same form of potential in the Einstein frame with the $R+R^2$ gravity (see \cite{BezrukovR2Higgs} with the interesting discussion
about differences in these two theories during post-inflationary dynamics). We see now that this connection exists also
on the level of Jordan frames. The $H \sim t$ dynamics represents the Ruzmaikin solution in quadratic gravity known for decades.
It is unstable for the $R+R^2$ theory which is used in the Starobinsky inflation scenario~\cite{Starobinsky}.  In the present subsection we have found an asymptotic solution with the same time behavior of $H(t)$ in the Jordan frame of nonminimally coupled scalar field theory
with $n=2N$.

It is interesting that in the $N=2$ case we can construct a polynomial potential for which a solution with the Hubble parameter evolving
linearly in time becomes an {\it exact} solution.  Indeed, let us assume that
\begin{equation}
\label{HR}
H=H_0t+C_0,
\end{equation}
where $C_0=-H_0t_0$.

It is easy to see that for $B(\varphi)=\varphi^2$ Eq.~(\ref{Fr21}) has solution (\ref{HR}) at
\begin{equation}
\varphi=\frac{H_0t+C_0}{\sqrt{3H_0(2\xi-1)}}.
\end{equation}
Then, from (\ref{equ00}) we get that
\begin{equation}
V(\varphi)=H_0\left(9(1-2\xi)\xi\varphi^4-\frac{3(1+2\xi)}{2}\varphi^2-\frac{1}{6(2\xi-1)}\right).
\end{equation}
To get real $\varphi(t)$ at $\xi<0$ we should put $H_0<0$. If $C_0>0$, then the inflationary scenario can be realized. At the first moment, $t=0$, $H=C_0$ and then tends to zero.
We do not consider such a potential which is not positively defined further in this paper.
Similar exact solutions have been found in the nonlocal gravity model~\cite{KVSQS2011}
(see also~\cite{KV2014}).

\section{Phase-space analysis models with power-law potentials}
\label{SectionPhP}
In this section, we investigate the general cosmological behavior of solutions by adopting a phase-space analysis.
The phase-space analysis allows the unfolding of many feature of the cosmological models with scalar fields~\cite{CapozzelloPhasespaceview,FaraoniDS}.
For the part of the obtained solutions the values of the scalar field and its derivative can go to infinity.
For this reason we introduce auxiliary variables, which are always restricted by the unit circulus. These variables have the following form
\begin{equation}
\alpha=\frac{\varphi}{\sqrt{1+{\varphi}^2+{\dot{\varphi}}^2}},
~~~~\beta=\frac{\dot{\varphi}}{\sqrt{1+{\varphi}^2+{\dot{\varphi}}^2}}.
\end{equation}
and the inverse transform formulae are
\begin{equation}
\label{alphabeta}
\varphi=\frac{\alpha}{\sqrt{1-{\alpha}^2-{\beta}^2}},
\qquad\dot{\varphi}=\frac{\beta}{\sqrt{1-{\alpha}^2-{\beta}^2}}.
\end{equation}
    Differentiating $\alpha$, $\beta$ and using (\ref{alphabeta}), we get
\begin{equation}
\label{absystem}
\begin{split}
\dot{\alpha}&=\beta\left(1-\alpha^2-\alpha\ddot{\varphi}\sqrt{1-{\alpha}^2-{\beta}^2}\right),
\\
\dot{\beta}&=\ddot{\varphi}\sqrt{1-{\alpha}^2-{\beta}^2}(1-{\beta}^2)-\alpha{\beta}^2.
\end{split}
\end{equation}

\begin{figure}[!h]
\centering
\includegraphics[width=72mm]{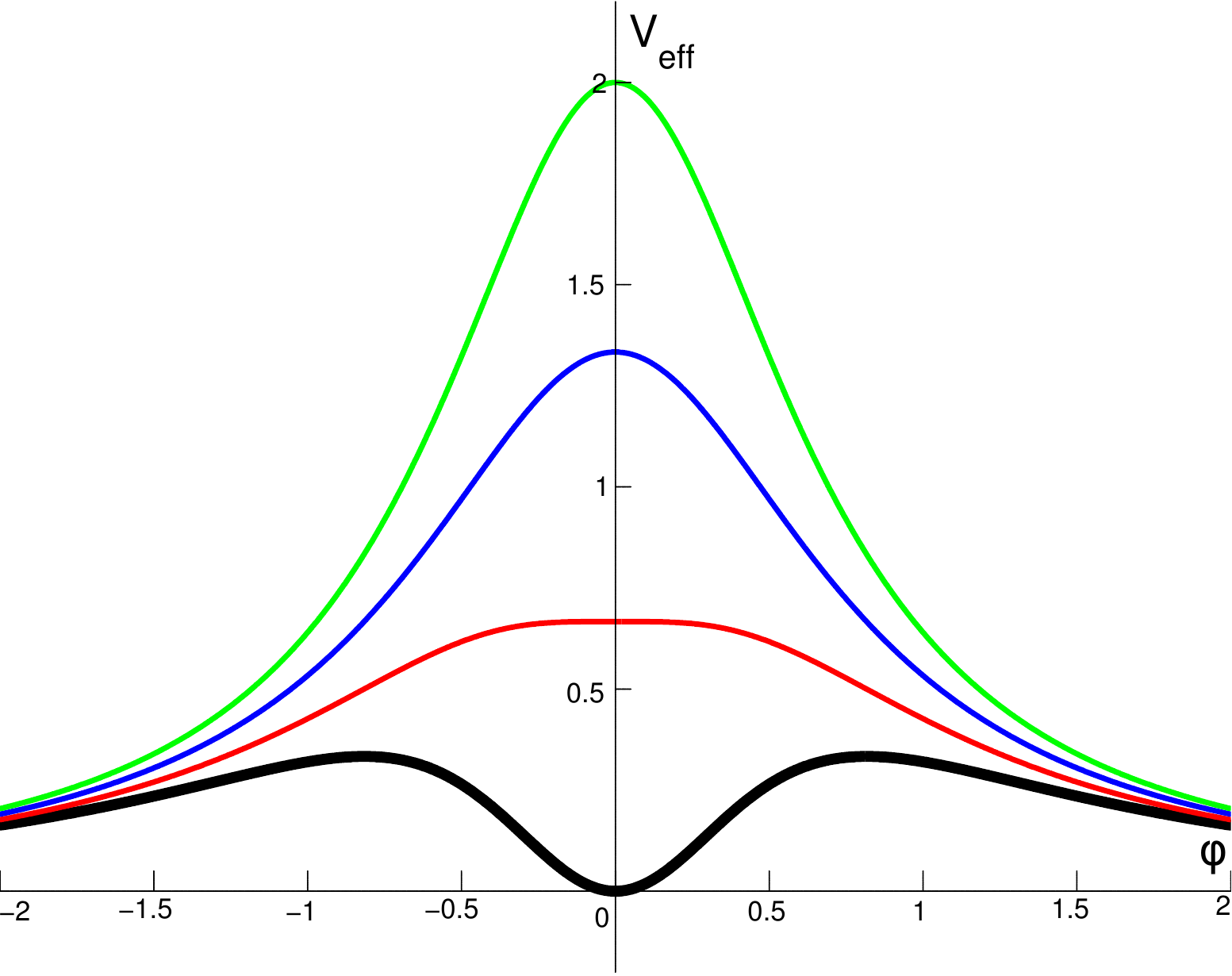} \ \ \ \ \ \  \  \  \
\includegraphics[width=61mm]{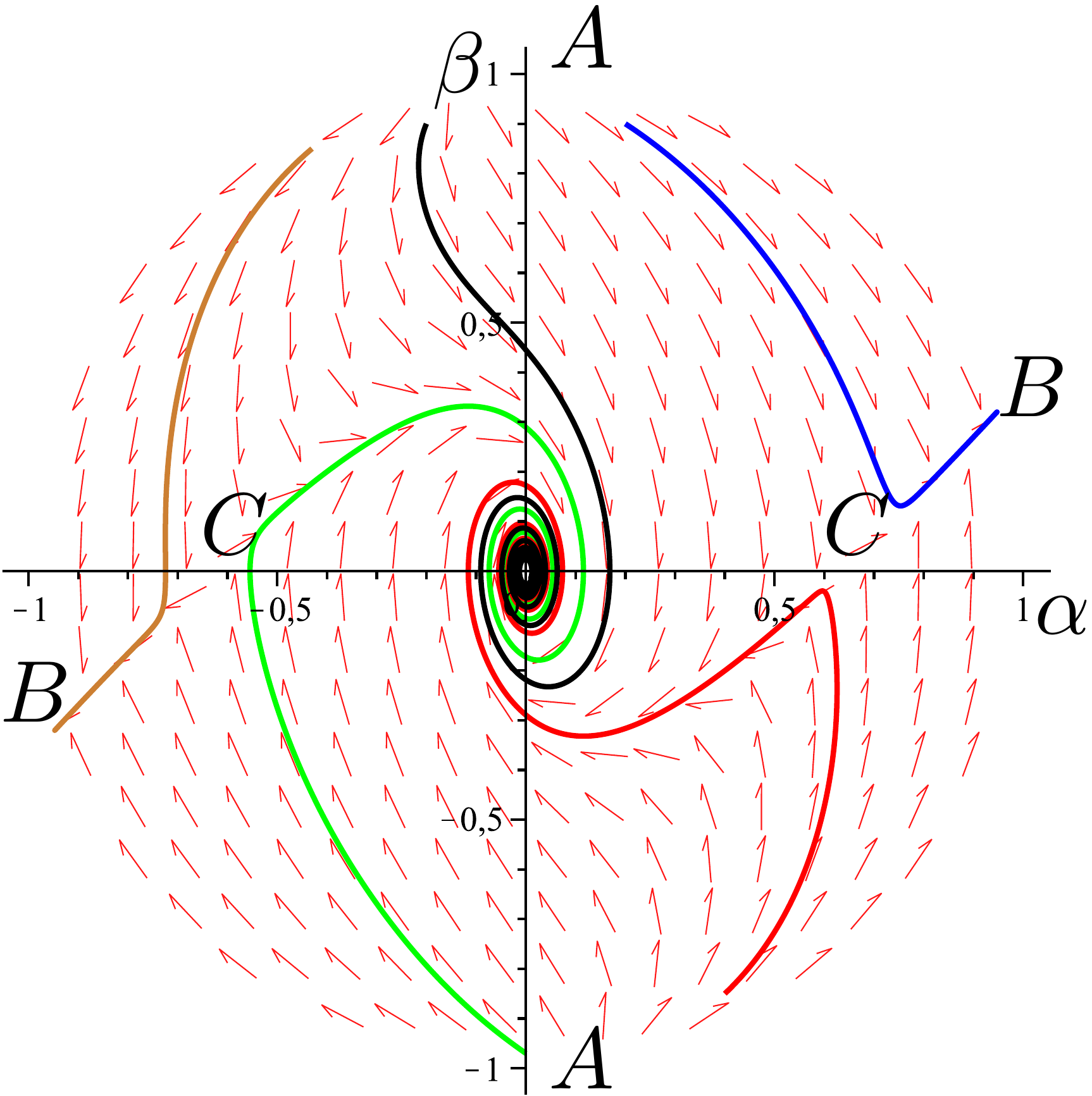}
\caption{The effective potentials $V_{eff}=\frac{V_0\varphi^2+\Lambda}{{(1-6\xi\varphi^2)}^2}$ for $N=2$ and $n=2$ (left) and the corresponding
phase portrait at $\Lambda=0$.
In both pictures $V_0=2$, $\xi=-\frac{1}{4}$. In the left picture  $\Lambda=0$ (black curve);  $2/3$ (red curve);   $4/3$ (blue curve); $2$ (green curve). In the right picture points $A$ correspond to power-law solutions (\ref{N2sol1}), points $B$ are exponential solutions (\ref{18}), and points $C$ are de Sitter solutions (\ref{dSSol}).}
\label{EPV2B2}
\end{figure}

We start with the case of $N=n$, considering a "classical" action with $N=n=2$ studied
many times in a lot of papers. The form of the effective potential (see black curve in Fig.~\ref{EPV2B2} (left plot))  shows clearly a possibility of two different late-time asymptotics. One of which is represented by scalar
field oscillations near the minimum of the potential, and the second corresponds to infinite growth
of the scalar field.

The typical phase diagram in the variables $\alpha$ and $\beta$  is shown in Fig.~\ref{EPV2B2} (right plot).
The solution (\ref{N2sol1}) corresponds to source points $A$.
The exponential solution  (\ref{18}) corresponds to points $B$ at the limiting circle. It attracts  part
of the trajectories, and the rest of them are attracted by the focus in the point $(0,0)$, which
represents scalar field oscillations. The unstable de Sitter solutions (\ref{dSSol}) are located in the
points $C=(\pm\sqrt{-1/(6\xi)},0)$ and separate these two stable regimes. The picture is qualitatively the same
for any negative $\xi$. It should be noted that the case of $N=2$ is a special one, and the form of asymptotic regimes
given above shows this clearly:  power indices of corresponding power-law solutions depend on $\xi$ for $N=2$
and do not depend on the coupling constant for other $N$. However, we have checked that for negative $\xi$
the form of the phase-space diagram does not depend on $\xi$  for $N=2$ also. We leave the more complicated case of $\xi>0$ for
a future work.

\begin{figure}[!h]
\centering
\includegraphics[width=76mm]{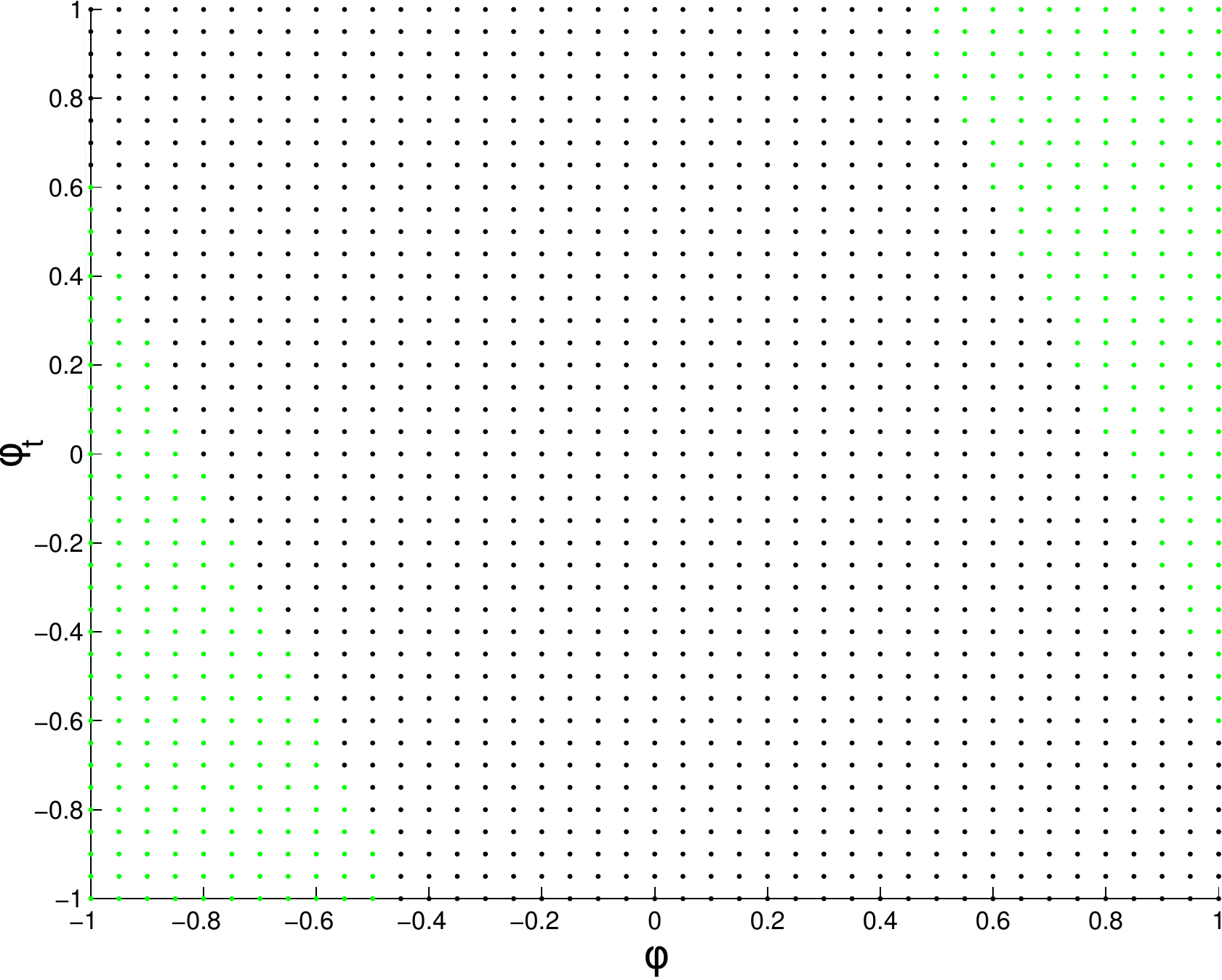}\  \  \  \ \ \
\includegraphics[width=76mm]{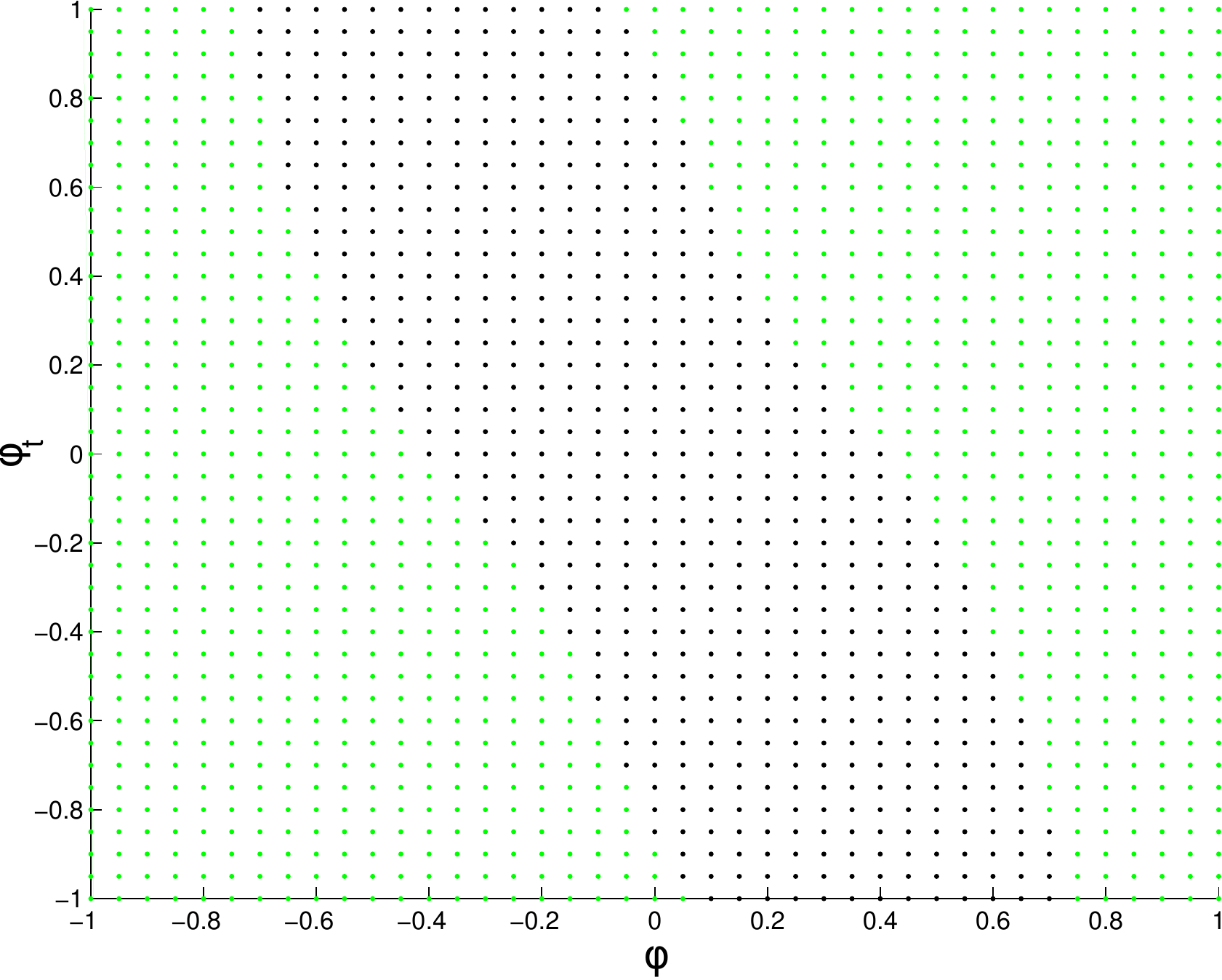}
\caption{The basin of attraction of the oscillations ( black points) and the exponential solution (\ref{18}) (green points) for $V=\varphi^2$, $B=\varphi^2$. We take $\xi=-\frac{1}{4}$ (left), $\xi=-1$ (right).}
\label{Bas}
\end{figure}

Going back to initial variables, it is possible to construct basins of attraction of these
two asymptotic regimes. We plot it in the variables $(\varphi, \dot \varphi)$, expressing $H$ through
the constraint equation. In general, there are two solutions for $H$. It is easy to see that
for $\xi<0$ and the positively determined potential
the minus sign  in Eq.~(\ref{H}) leads to initial $H<0$, so in order to study an expanding universe
we fix the sign to be plus. After that, the point $(\varphi, \dot \varphi)$ fixes the initial data
completely. Figure~\ref{Bas} shows that the basin of attraction of the oscillations is located inside a  band in
the initial conditions space. Two boundary points of this band at $\dot \varphi=0$ evidently correspond to two maxima
of the effective potential. A nonzero initial velocity could result in overcoming the potential hill when starting in its direction somewhere from the "bottom", so the boundary appears to be inclined with respect to the coordinate axis. The band becomes narrower with
increasing $|\xi|$.

\begin{figure}[!h]
\centering
\includegraphics[width=60mm]{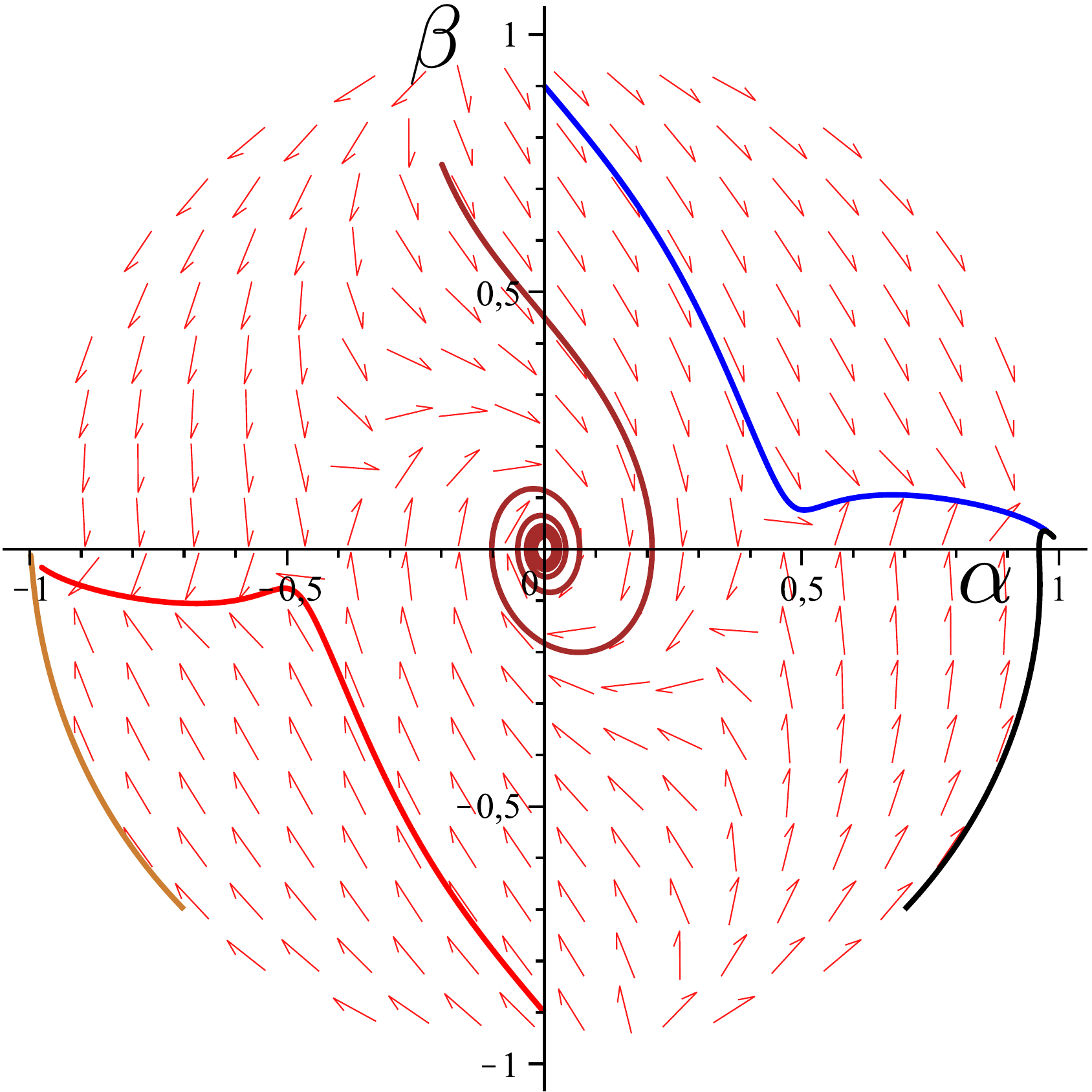} \ \ \ \ \ \ \
 \includegraphics[width=72mm]{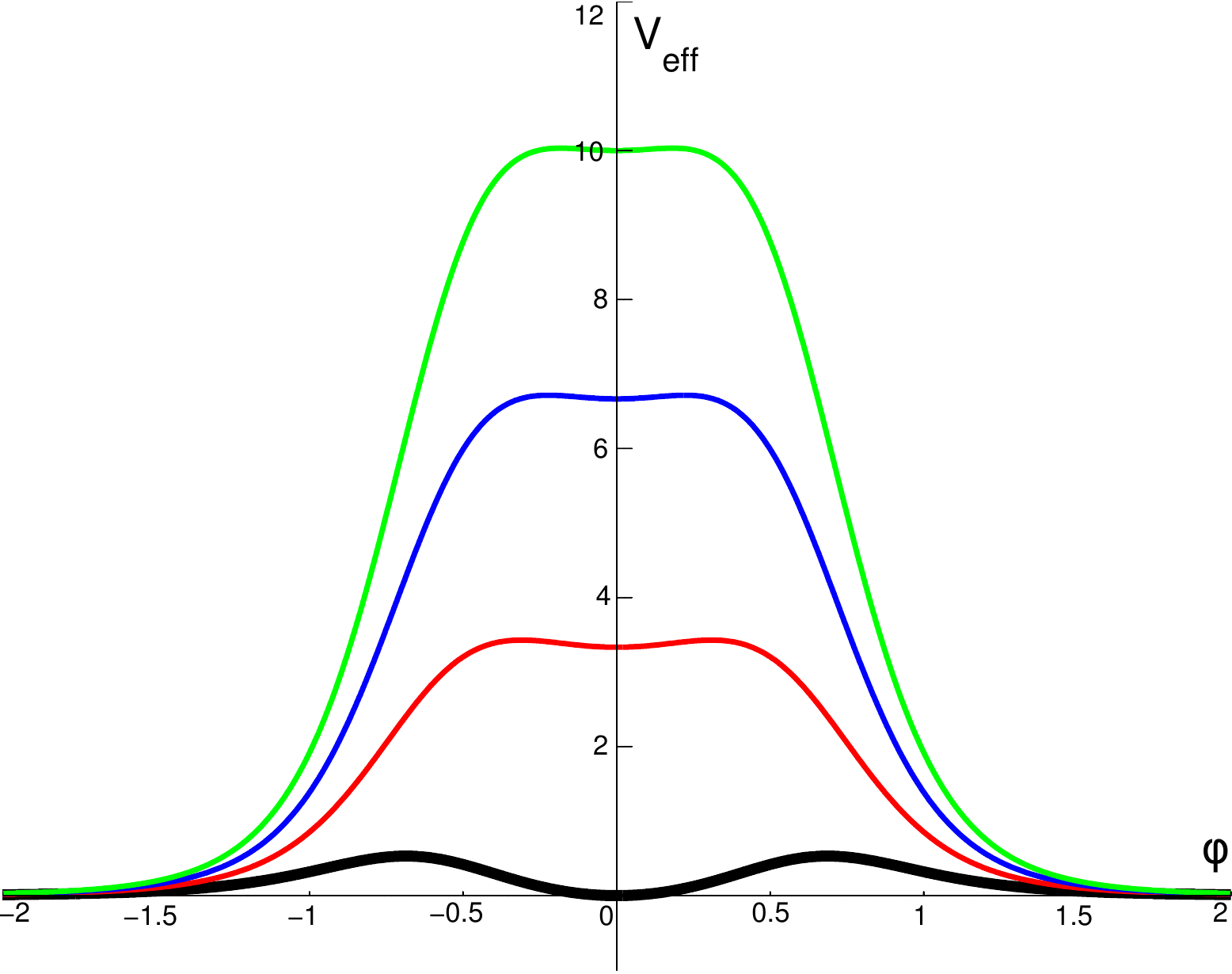}
\caption{The solution of system (\ref{absystem}) with $V=V_0\varphi^2$ and $B=\varphi^4$. We put $\xi=-1$, $V_0=1$, and  $\Lambda=0$ (left).
The effective potential $V_{eff}=\frac{V_0\varphi^2+\Lambda}{{(1-6\xi\varphi^4)}^2}$ for $V=V_0\varphi^2+\Lambda$ and $B=\varphi^4$, $\xi=-1/4$, $V_0=4$ and $\Lambda=0$ (black curve);~ $10/3$ (red curve);~  $20/3$ (blue curve);~ $10$ (green curve).}
\label{V2B4}
\end{figure}

If $n<N$ the general picture of dynamics (Fig.~\ref{V2B4}) is similar to the above case. The only difference
is that the exponential stable solution is replaced by power-law regime (\ref{powerlawN}). Correspondingly,
the location of this stable point at the limiting circle in the $(\alpha, \beta)$ plane is shifted to the points
$(\pm 1, 0)$.

\begin{figure}[!h]
\centering
\includegraphics[width=72mm]{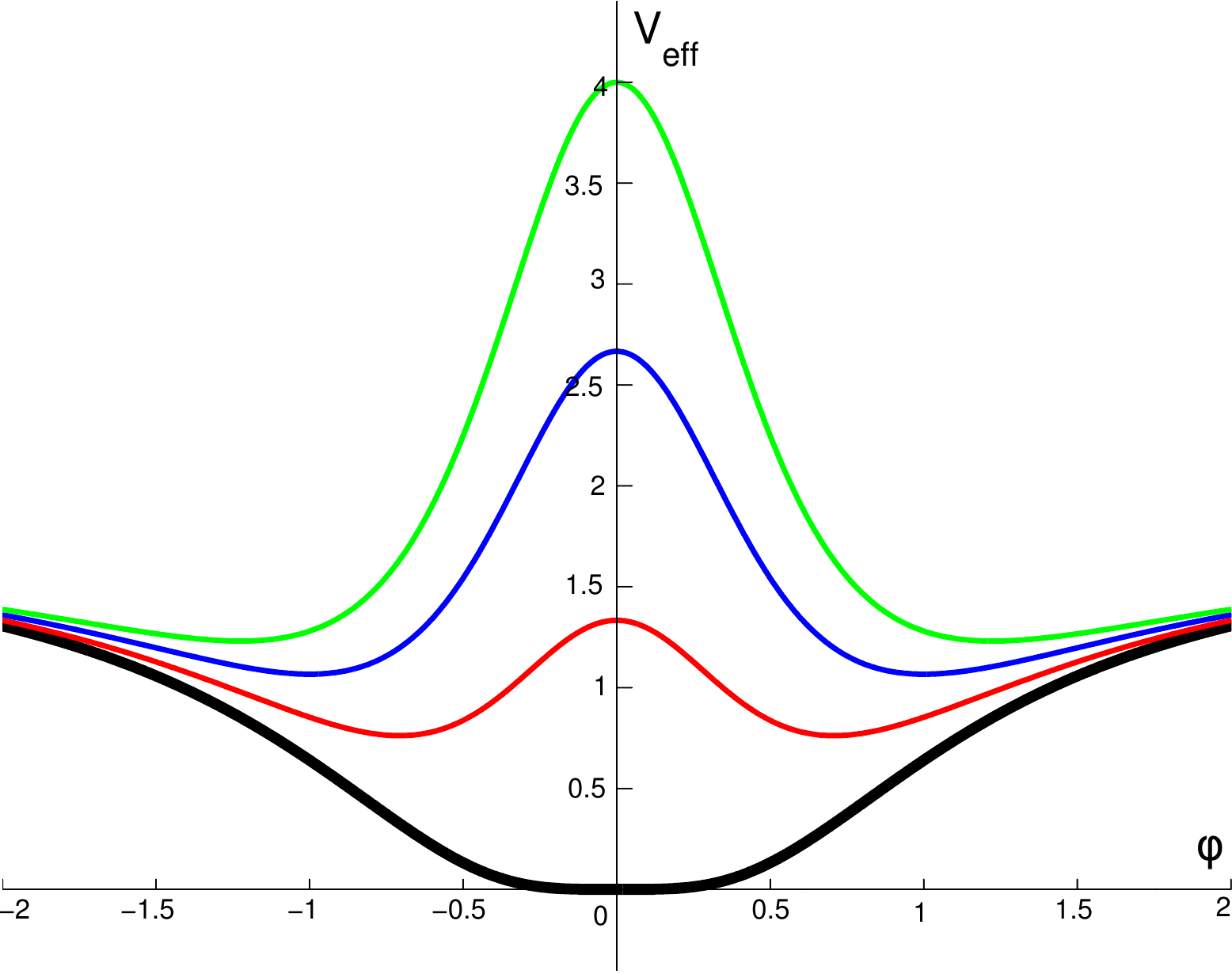}
\caption{The effective potential $V_{eff}=\frac{V_0\varphi^n+\Lambda}{{(1-6\xi\varphi^N)}^2}$ for $N=2$,$ n=4$, $\xi=-\frac{1}{4}$, $V_0=4$ and $\Lambda=0$ (black curve);~ $4/3$ (red curve);~  $8/3$ (blue curve);~$4$ (green curve).}
\label{VeffV4B2}
\end{figure}

\begin{figure}[!h]
\centering
\includegraphics[width=72mm]{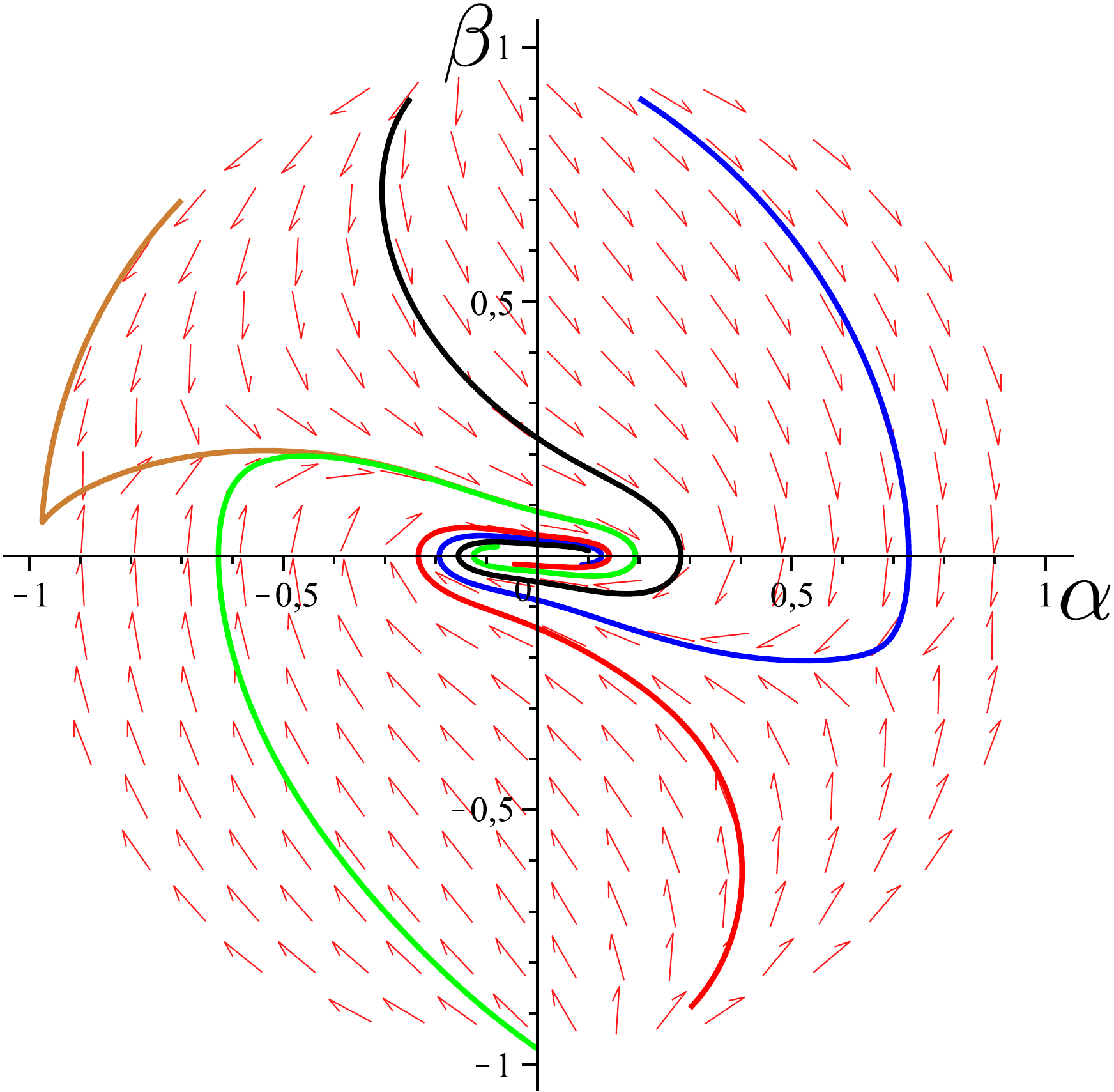} \  \  \  \  \ \ \ \ \ \
\includegraphics[width=72mm]{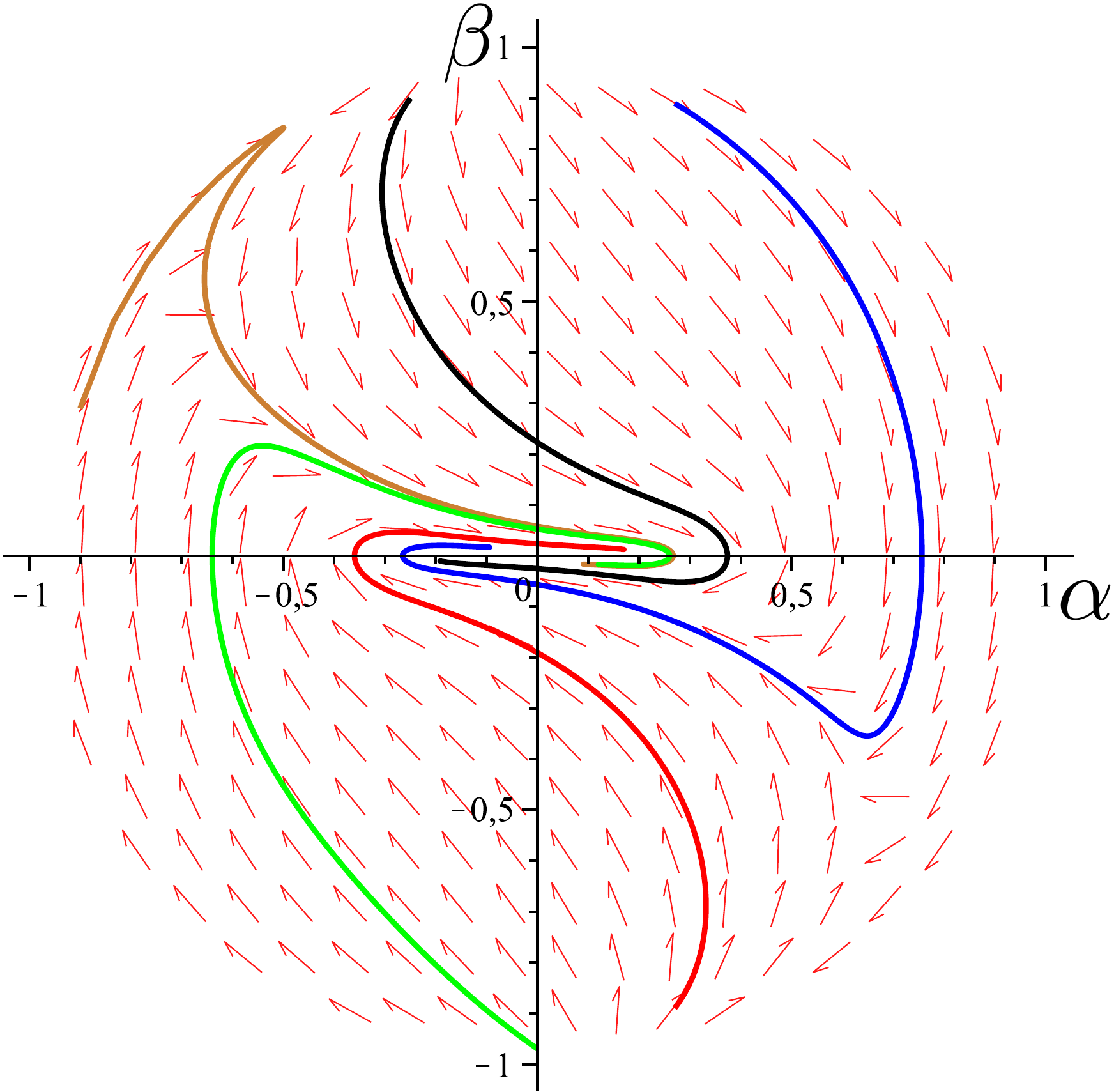}
\caption{The phase portrait of system (\ref{absystem}) with $B=\varphi^2$,  $\xi=-1/4$ and
$\Lambda=0$. The potentials are  $V=\varphi^4$ (left), $V=\varphi^6$ (right). }
\label{Lambda0}
\end{figure}

For the case of $n=2N$ there are no other
attractors different from oscillations (see the corresponding effective potential
in Fig.~\ref{VeffV4B2}, black  curve), so all trajectories tend to the $(0,0)$ point. We present the phase portrait for $N=2$ and $n=4$ in Fig.~\ref{Lambda0} (left). Correspondingly the unstable
de Sitter solution is absent. The Rusmaikin-type solution appears as an unstable fixed point
at $(\pm 1, 0)$.

\begin{figure}[!h]
\centering
\includegraphics[width=72mm]{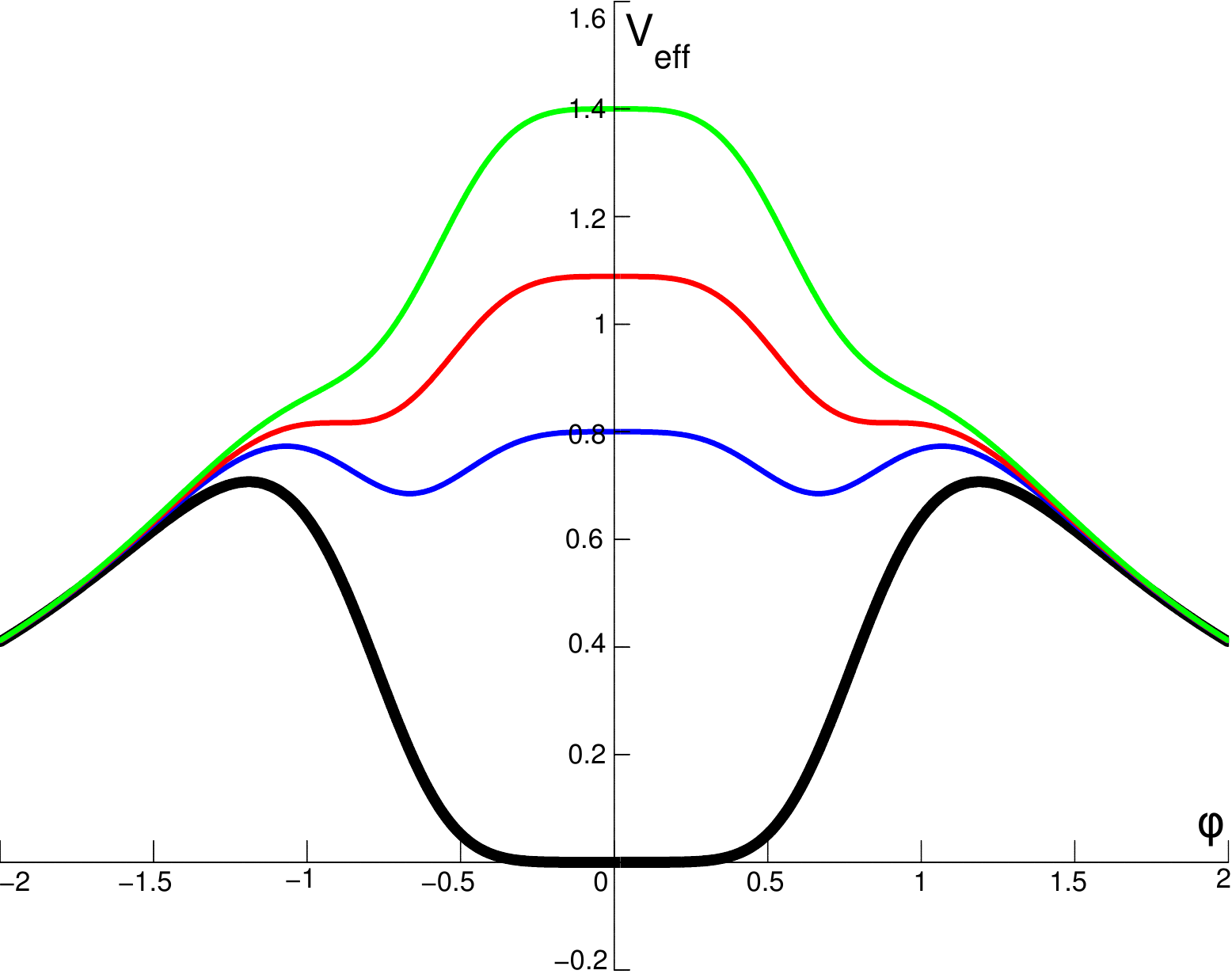} \  \  \  \  \ \ \ \ \ \
\includegraphics[width=60mm]{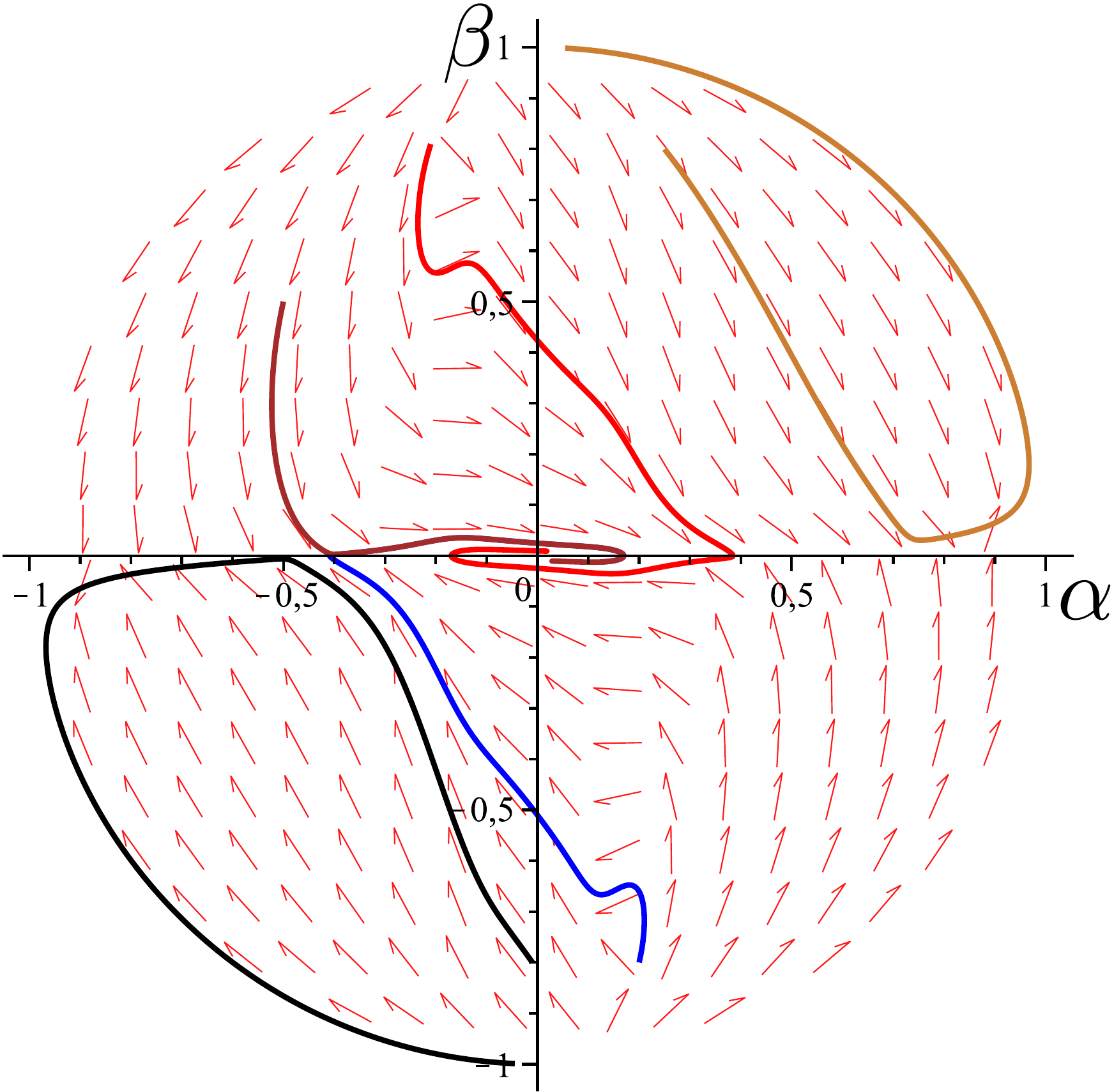}
\caption{In left picture there is the effective potential $V_{eff}=\frac{V_0\varphi^n+\Lambda}{{(1-6\xi\varphi^N)}^2}$ for $N=4$, $n=6$, $\xi={}-1/4$, $V_0=4$ and $\Lambda=0$ (black curve); ~ $0.8$ (blue curve);~  $1.0887$ (red curve);~  $1.4$ (green curve). In right picture there is the phase portrait of system (\ref{absystem}) with $V=5\varphi^6$, $B=\varphi^4$,  $\xi={}-10$. }
\label{V6B4}
\end{figure}

The situation for $N<n<5N$, $n \ne 2N$ has its own difficulties. It is the only case when
the set of variables $(\alpha, \beta)$ is not good enough for presenting numerical results
because two different points -- unstable (\ref{9}) and power-law solution  (\ref{powerlawN}) are projected at the same
point $\alpha=0, \beta=1$. For $n>2N$ this does not lead to big problems,
because the  power-law
solution (\ref{powerlawN}) (which is present in Fig.~\ref{Lambda0} (right) and is absent in Fig.~\ref{Lambda0} (left) is unstable and its location coincides with the unstable point (\ref{9}).
There is only one future asymptotic in the form of scalar field oscillations.

As it is mentioned in the previous section, for very steep potentials ($n>5N$, $N \ne 2$ or $n>10$, $N=2$, $\xi<\xi_{cr}$) the saddle
disappears and the point $(0,1)$ restores its simple node nature. We do not consider such steep potentials here.

The last case is $N<n<2N$ where a stable Big Rip solution exists. Several typical trajectories have been shown in Fig.~\ref{V6B4}, two of them start
and finish at the same point on the plane $(\alpha, \beta)$  [the points $(0, \pm 1)$]. However, it interpolates between
two different asymptotic solutions given by (\ref{9})  in the past  and (\ref{powerlawN}) in the future in the original variables.

\section{More complicated potentials}

\label{SectionPolyPot}
\begin{figure}[!h]
\centering
\includegraphics[width=72mm]{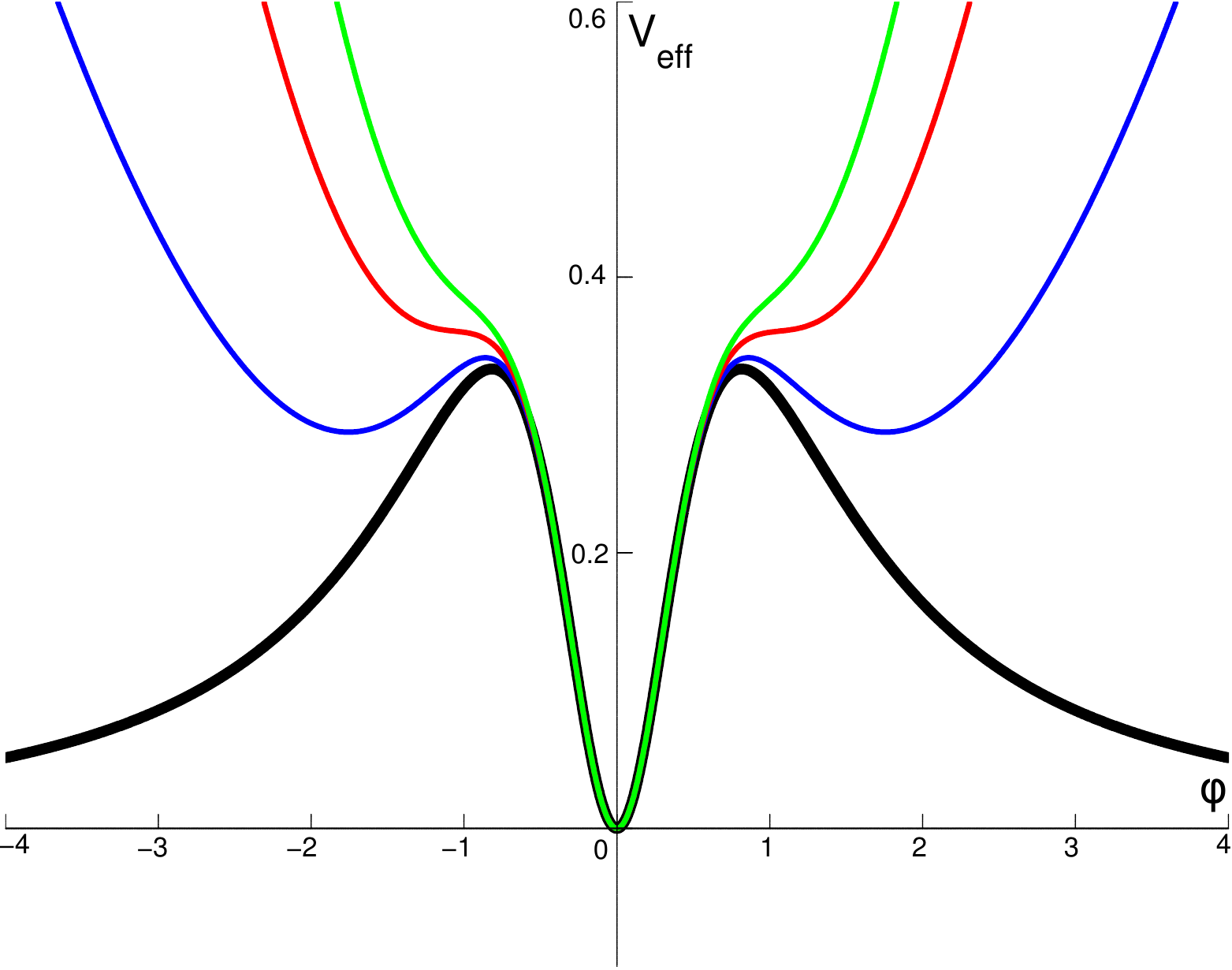} \ \ \ \ \ \ \ \ \ \ \
\includegraphics[width=60mm]{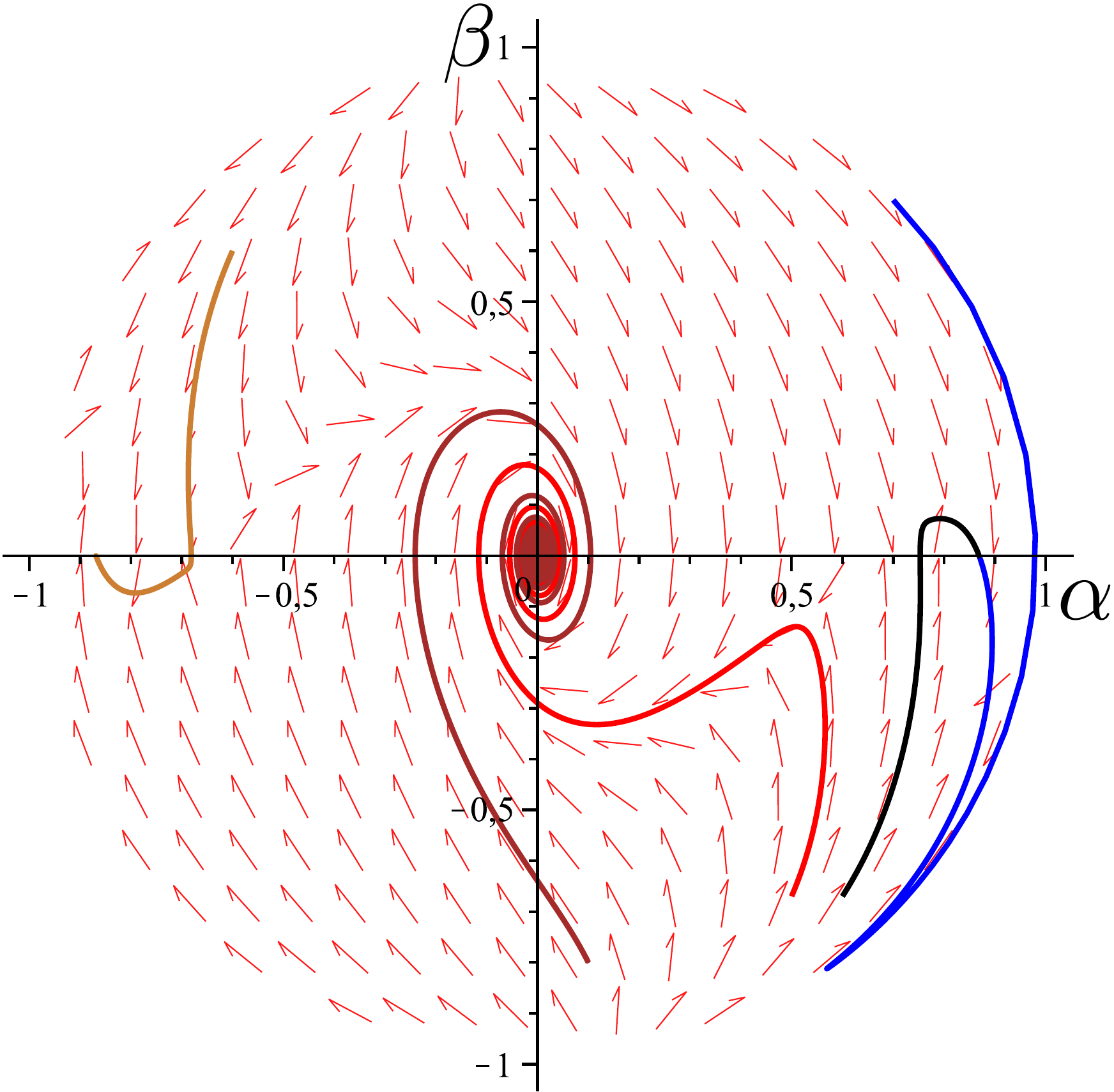}
\caption{The effective potential $V_{eff}=\frac{V_0\varphi^2+\lambda\varphi^6}{{(1-6\xi\varphi^N)}^2}$ for $N=2$, $\xi=-\frac{1}{4}$, $V_0=2$ and $\lambda=0$ (black curve); ~ $0.1$ (blue curve);~ $0.25$ (red curve);~ $0.4$ (green curve) and the phase portrait of system (\ref{absystem}) with the corresponding potential $V$ at $\lambda=0.1$.}
\label{V6_2}
\end{figure}

If the scalar field potential is more complicated than a single power-law term, the resulting
dynamical picture can be richer. As an example we mention here an important case of
$V= V_0 \varphi^2 + \lambda \varphi^n$, where $n>4$. For small enough $\lambda$ the
local maximum of the effective
potential continues to exist, however, a new local minimum appears.
This results in the disappearance of exponential solutions existing for $\lambda=0$ and  the appearance of a stable de Sitter solution
instead (note, that we have not added any explicit cosmological constant to  action (\ref{action}) in order to get this de Sitter solution).
Examples of an effective potential for the case of $V= V_0\varphi^2+\lambda \varphi^6$ and $N=2$ for different $\lambda$ are shown in Fig.~\ref{V6_2} (left).
The critical value of $\lambda_{cr}=4\sqrt{3}(2-\sqrt{3})\xi^2 V_0$ corresponds to the disappearance of the de Sitter solution which exists
for $\lambda<\lambda_{cr}$. At $\lambda>\lambda_{cr}$ the only future asymptotic is oscillation near the $(0,0)$ point.

\begin{figure}[!h]
\centering
\includegraphics[width=50mm]{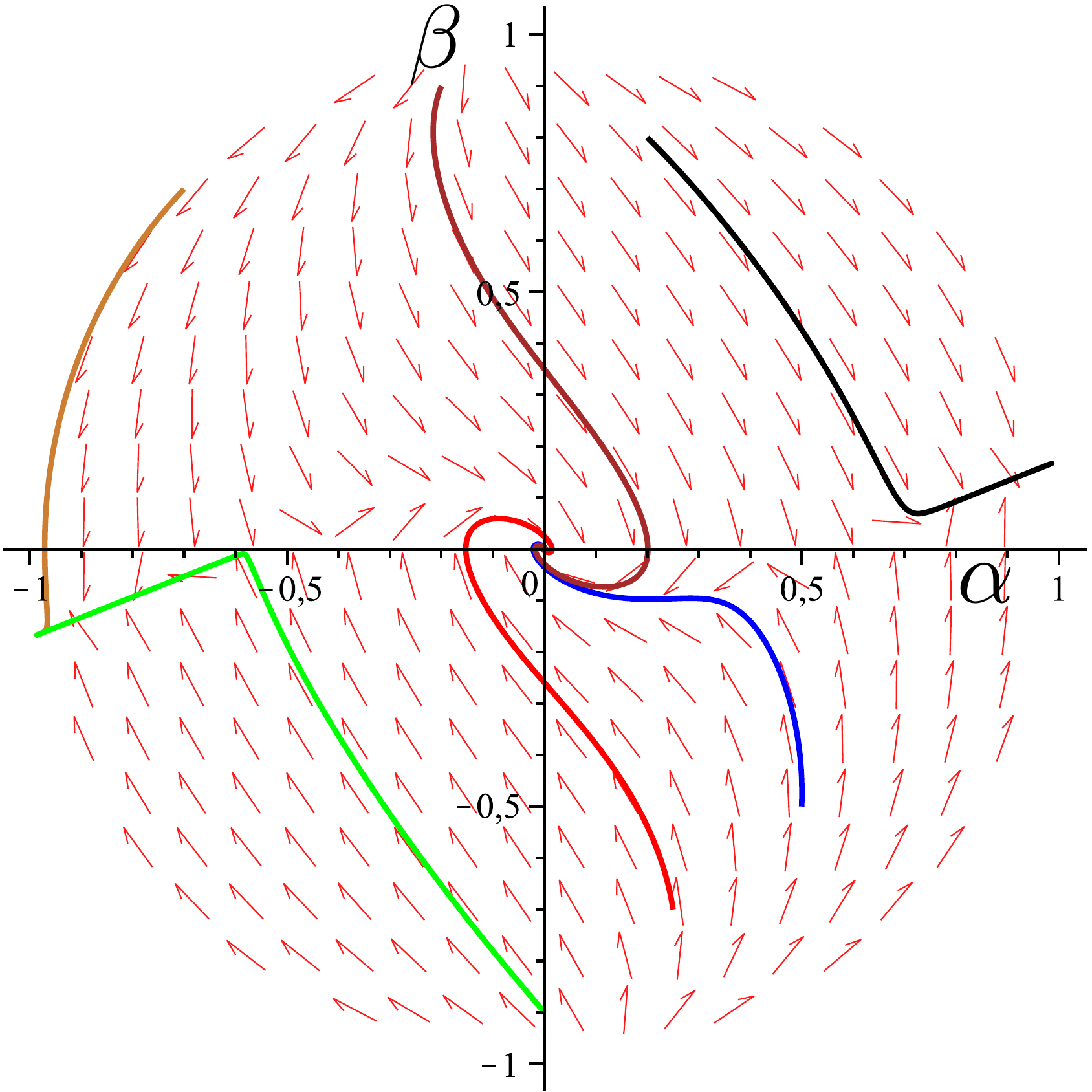} \  \  \  \
\includegraphics[width=50mm]{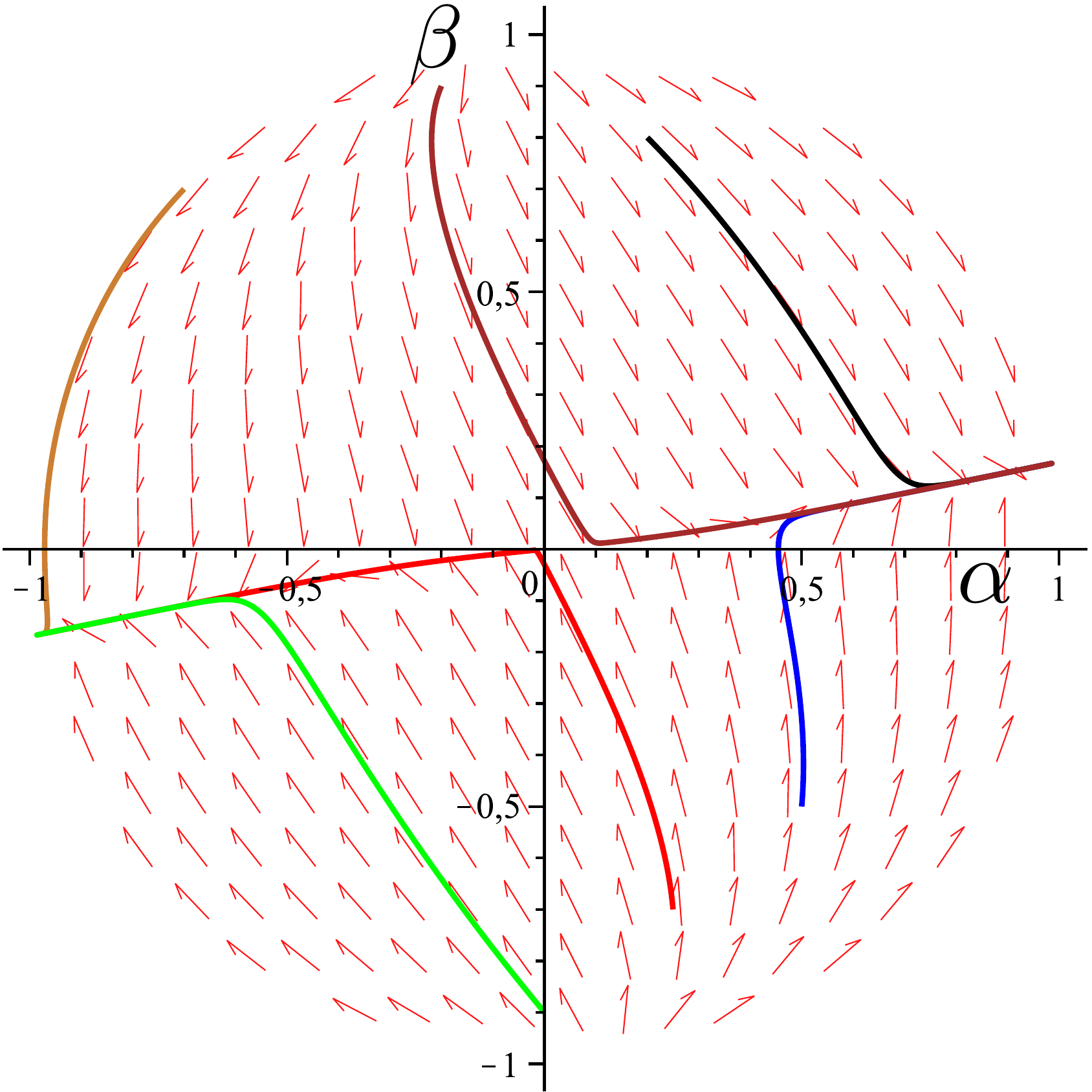} \  \  \  \
\includegraphics[width=50mm]{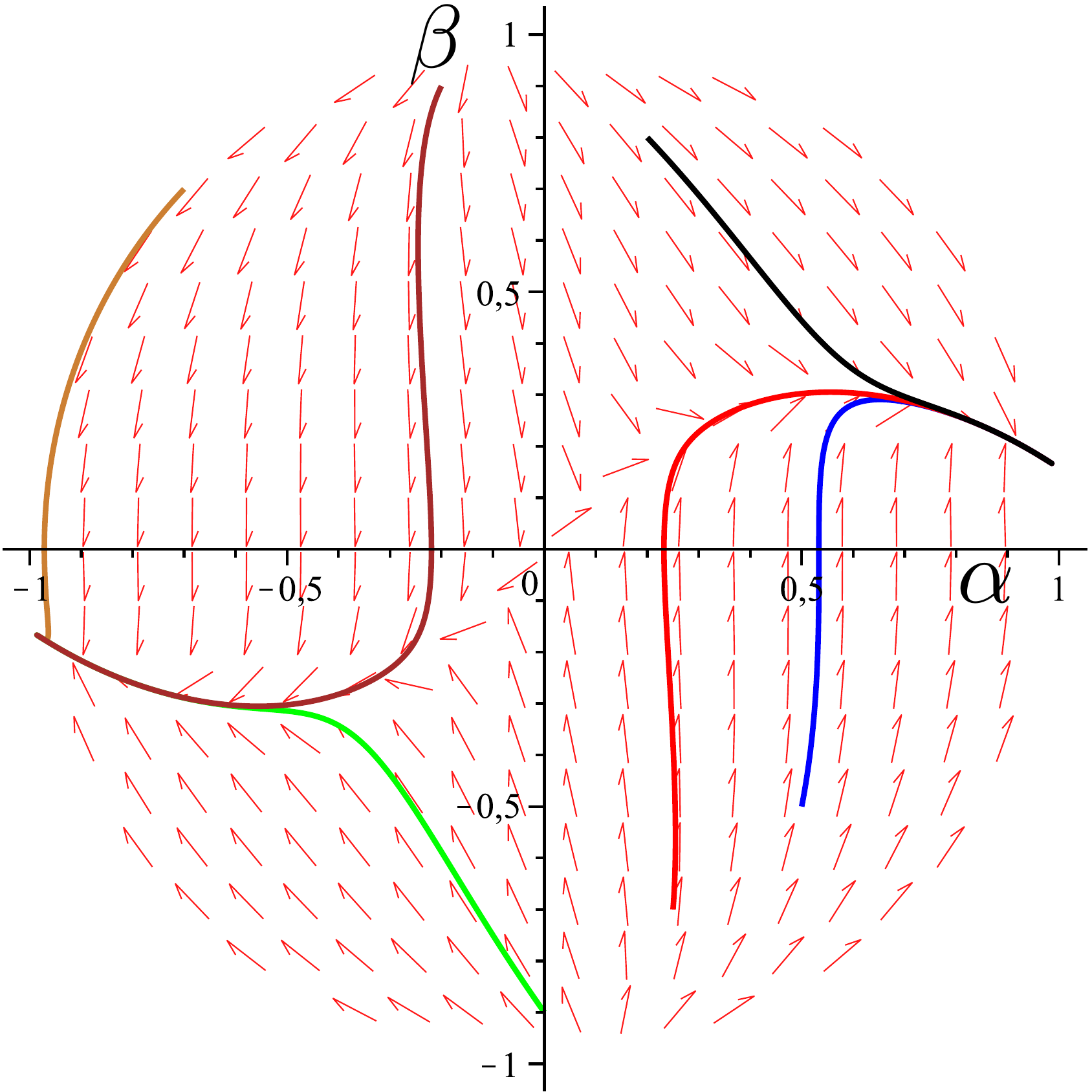}
\caption{The solution of system (\ref{absystem}) with $B=\varphi^2$ and $V=V_0\varphi^2+\Lambda$. We choose $\xi=-1/4$, $V_0=1/2$ and  $\Lambda=0.05$ (left), $\Lambda=0.2$ (middle) or  $\Lambda=1$ (right).  }
\label{V2B2}
\end{figure}

The simplest way to get stable de Sitter solutions is to add a positive constant ($\Lambda$ term) to the potential, however, the resulting
picture is not as simple as in the case of minimally coupled field.  Let us consider the potentials $V=V_0\varphi^n+\Lambda$, with a constant $\Lambda>0$.
It is easy to see from the form of the effective potential that for the case of $N=n$ (see Fig.~\ref{EPV2B2}) a large enough positive cosmological constant destroys the quasi-Einstein
regime of scalar field oscillations, and all trajectories tend to corresponding
stable  exponential solution (see Fig.~\ref{V2B2}).
 In the case of $n=2$ and $N=2$ the effective potentials for different $\Lambda$ are present in Fig.~\ref{EPV2B2}.
 The condition for the oscillatory solution to exist can easily be derived from the form of the effective potential.
  At $\Lambda<-V_0/(12\xi)$ there exists a stable de Sitter solution at $\varphi=0$ and two unstable de Sitter solutions at
 \begin{equation*}
 \varphi_{dS} = \pm\frac{\sqrt{6V_0(V_0+12\xi\Lambda)}}{6V_0\sqrt{-\xi}}.
 \end{equation*}
 At large $\Lambda$ the model has an unstable de Sitter solution at $\varphi=0$ only.

It is interesting that though a run-away asymptotic exists also for $n<N$, the basin of attraction
of the oscillatory solution, though shrinking, continues to exist for an arbitrary large $\Lambda$.  It can be shown
analytically using the fact
that the effective potential has a minimum at the point $\varphi=0$ for any $\Lambda \geqslant 0$. In particular,
  at $n=2$ we get
\begin{equation}
\label{V2effLambda}
    V''_{eff}(0)=2V_0+24\xi\Lambda\delta_{2,N},
\end{equation}
where $\delta$ is  Kronecker's symbol, so $V''_{eff}(0)>0$ for $N>2$. The case $N=4$ is presented on Fig.~\ref{V2B4}.

\begin{figure}[!h]
\centering
\includegraphics[width=72mm]{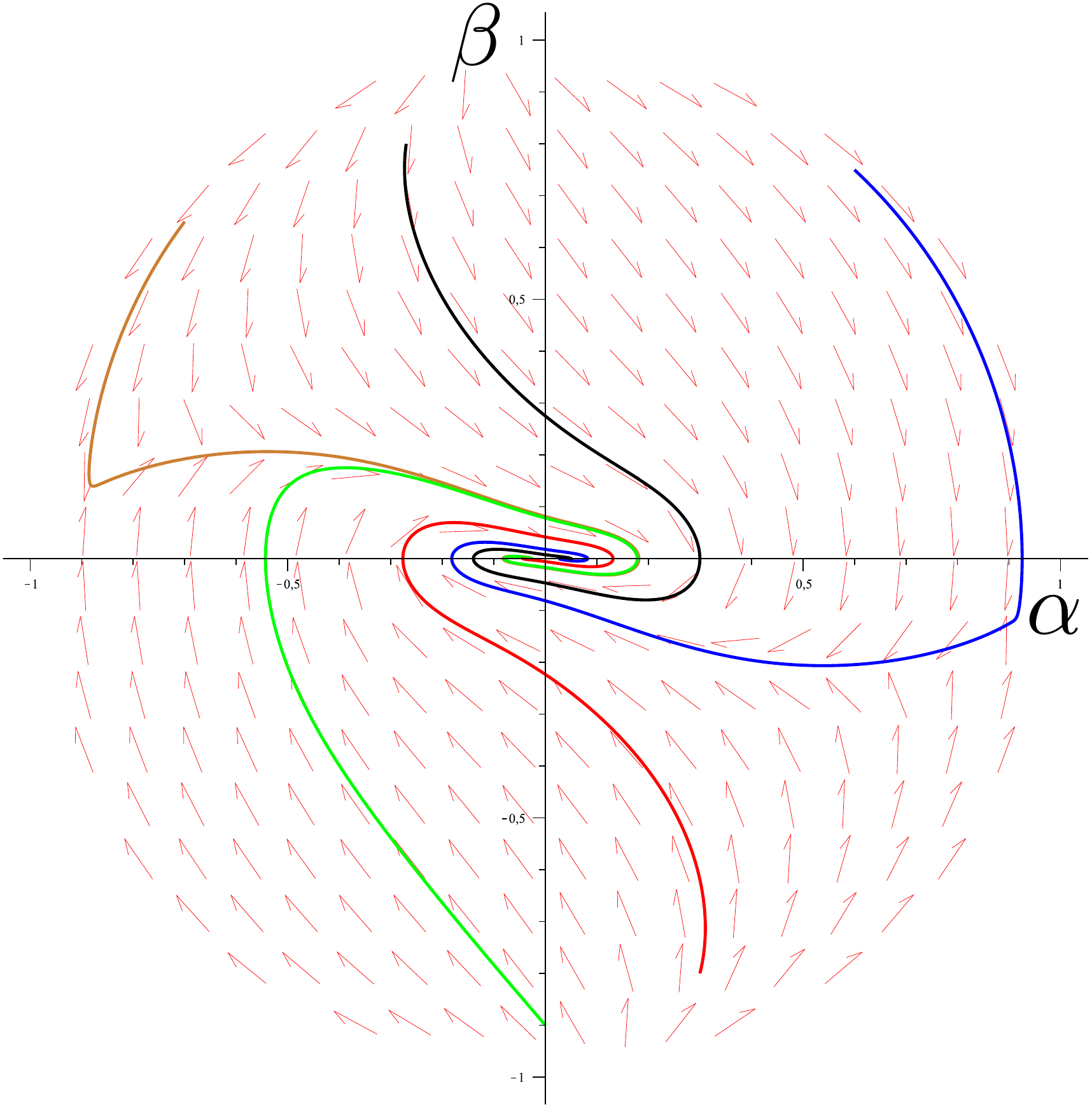} \ \ \ \
\includegraphics[width=72mm]{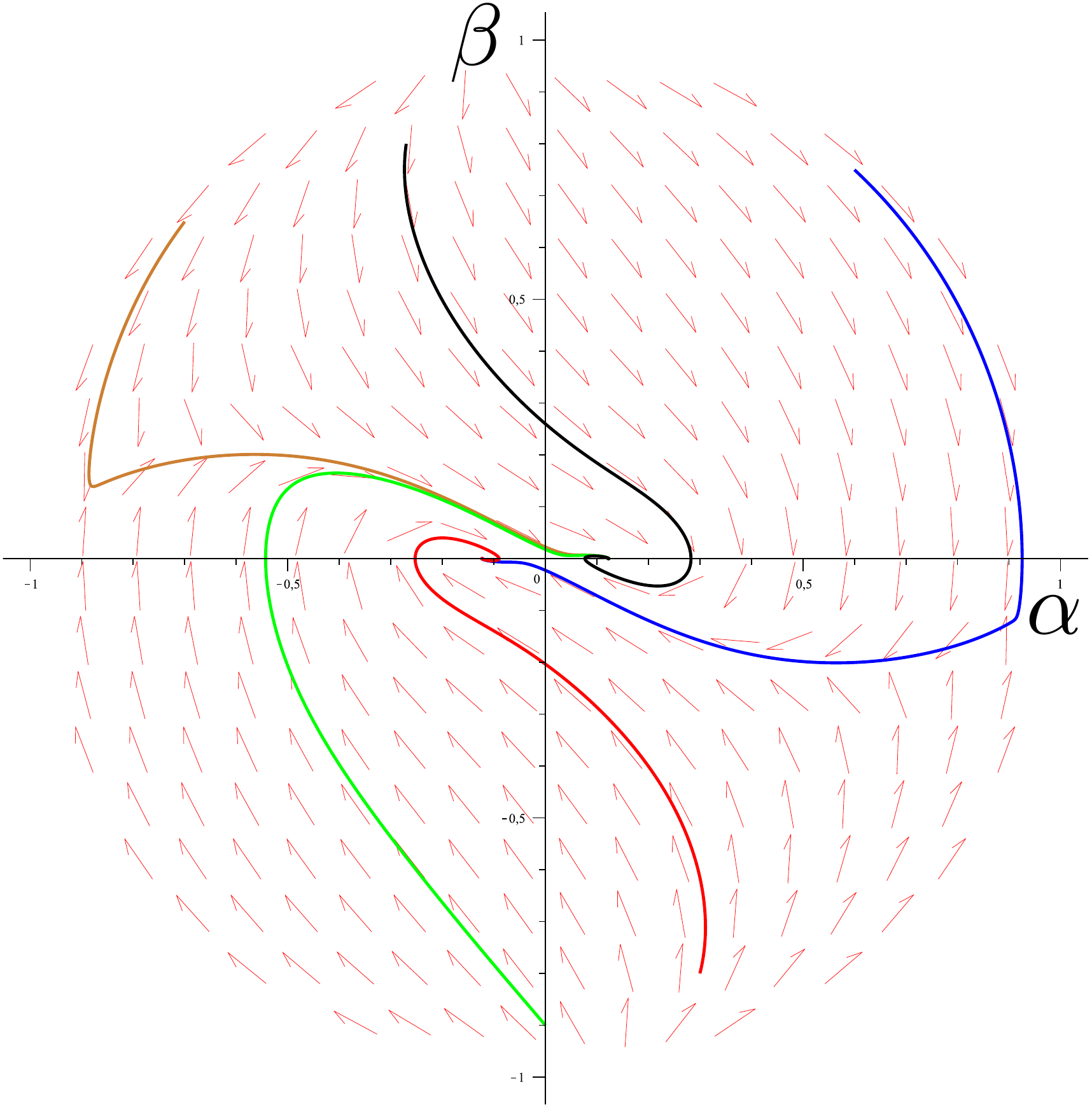} \\[2.7mm]
\includegraphics[width=72mm]{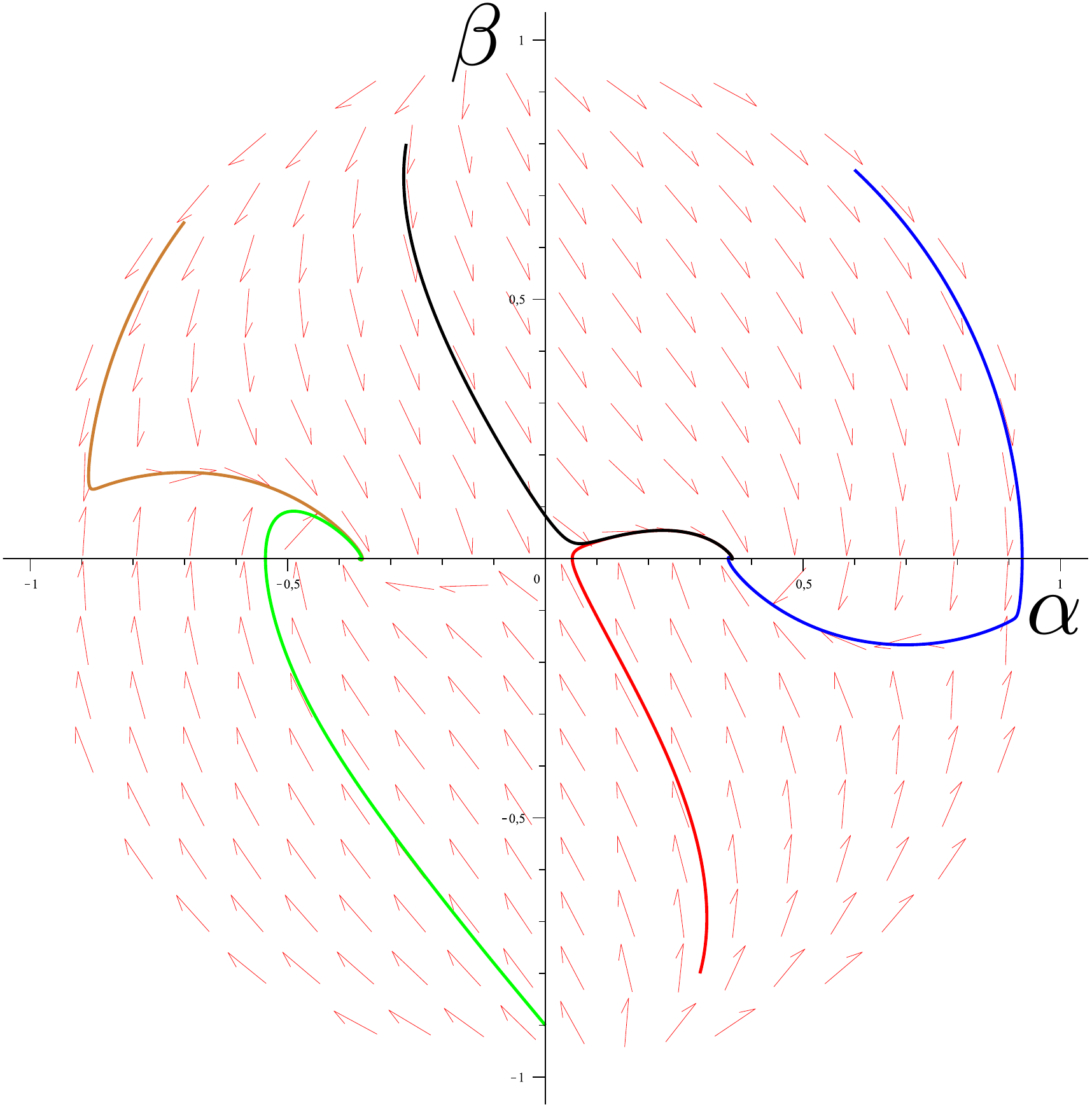} \ \ \ \
\includegraphics[width=72mm]{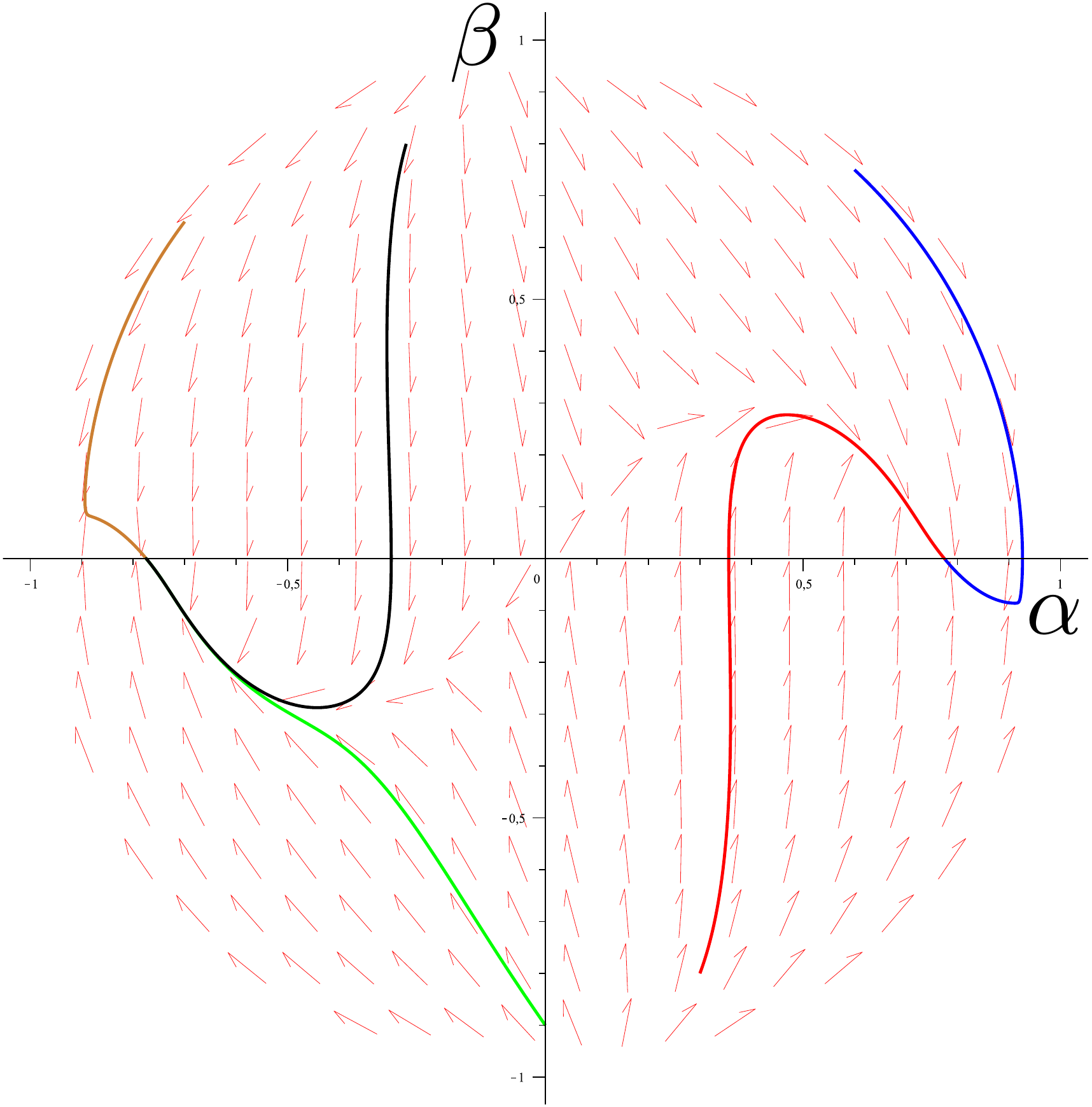}
\caption{The solution of system (\ref{absystem}) with $V=\varphi^4+\Lambda$, $B=\varphi^2$, $\xi=-1/4$.
 We choose $\Lambda=0.001$, $\Lambda=0.01$,  $\Lambda=0.1$, and  $\Lambda=1$
  (from left to right). }
\label{V4B2}
\end{figure}

On the contrary, if $n>N$, then the point $(0,0)$
is a maximum for any positive $\Lambda$.
For example,
if $n>2$ and $N=2$, then
 \begin{equation}
\label{V2effLambdan}
    V''_{eff}(0)= 24\xi\Lambda<0,
\end{equation}
at any $\Lambda>0$.
Therefore, in these cases there are unstable de Sitter solutions at $\varphi=0$. Stable de Sitter solutions can exist for nonzero values of $\varphi$. For example, at $n=4$ we get  stable de Sitter solutions
\begin{equation*}
\varphi_{dS} =\pm \sqrt{\frac{-6\xi\Lambda}{V_0}},
\end{equation*}
at any nonzero value of $\Lambda$.
A similar situation occurs in general
for $n \geqslant 2N$ when the effective potential has minima independently of $\Lambda$.
 An example for $N=2$, $n=4$ is shown in Figs.~\ref{VeffV4B2} and \ref{V4B2}. For $n>2N$ the effective potential always has a Higgs-like form with two minima, and grows infinitely with $\varphi \to \infty$. So, the only possible future fate of a trajectory in the $n \geqslant 2N$ case with a cosmological constant is a stable de Sitter future asymptotic (see Fig.~\ref{V4B2}).

 Alternatively, big enough $\Lambda$ turns the effective potential to be monotonically
decreasing when $n<2N$, so the Big Rip alternative becomes the only possible one. This happens, for example, for $N=4$ and $n=6$
(see Fig.~\ref{V6B4}) if $\Lambda^2 > (V_0/2)^2 (1/6|\xi|)^3$ (the critical case corresponds to the red curve in Fig.~\ref{V6B4}).

Combining the results of this section we can see that the relation between the presence of the cosmological constant in the action of
the theory with nonminimal coupling and the existence of  a stable future de Sitter asymptotic is not direct (as in the case of minimal
coupling): there are theories with cosmological constant and no stable future de Sitter asymptotic ($V=\Lambda+ \varphi^n$ with
$N \leqslant n < 2N$ and $\Lambda$ bigger than the critical one) as well as theories with stable de Sitter without explicit $\Lambda$ in
the action ($V=V_0\varphi^n + \lambda \varphi^{n_1}$ where $n<2N$, $n_1>2N$, and small enough~$\lambda$).

\section{Conclusions}

In the present paper we have made a global qualitative analysis for the cosmological dynamics with a nonminimally
coupled scalar field with power-law coupling functions and potentials. Local analysis provided in \cite{Sami:2012uh} shows that for $\xi<0$
stability properties of asymptotic solutions do not depend on $\xi$ and depend only upon power indices $N$ and $n$ of the coupling function $B$ and
the scalar field potential $V$ correspondingly. This allows us to cover most interesting cases of small $N$ and $n$ by a limiting number of phase-space diagrams.

We argue that usage of effective potential $V_{eff}$ helps significantly to understand different cases of qualitatively different dynamics. This does not require a full transition to the Einstein frame which can be rather cumbersome.

We identify three qualitatively different points of scalar field dynamics realized for $n<N$, $N<n<2N$, and $n>2N$ and two boundary
cases of $n=N$ and $n=2N$. The latter is of particular interest because it contains (and generalizes) the Higgs inflation proposal.
All fixed points for corresponding dynamics found in \cite{Sami:2012uh} are incorporated into the global phase diagram constructed in the present paper.

Some interesting modifications of possible phase diagrams when more general potentials are allowed also have been presented.
In particular, we study the influence of the explicit cosmological constant in the action of the theory. It is interesting, that, unlike the
minimally coupled case, nonzero $\Lambda$ in the action does not automatically lead to the existence of a de Sitter solution.
The opposite is also true - there are theories (for example, with $V=V_0\varphi^2+\lambda \varphi^6$ and quadratic coupling) which have stable
de Sitter solutions without a cosmological constant in the action.

\medskip

\noindent {\bf Acknowledgements. \ }
The research of S.V. is supported in part by RFBR grant 14-01-00707 and by the Russian Ministry of Education and Science under grant NSh-3042.2014.2. A.T. is supported by RFBR grant 14-02-00894.
The work of A.T. is partially  supported by the Russian Government Program of Competitive Growth of Kazan Federal University.

\end{document}